\documentclass[preprint,prl,aps,floatfix,superscriptaddress]{revtex4-2}

\usepackage{graphicx}  % need for figures
\usepackage{bm} % bold math
\usepackage{times}

\begin{document}

\title{Observation of the Non-linear Meissner Effect}

\author{J. A. Wilcox}
\affiliation{H. H. Wills Physics Laboratory, University of Bristol, Tyndall Avenue, Bristol, BS8 1TL, United Kingdom}

\author{M. J. Grant}
\affiliation{H. H. Wills Physics Laboratory, University of Bristol, Tyndall Avenue, Bristol, BS8 1TL, United Kingdom}

\author{L. Malone}
\affiliation{H. H. Wills Physics Laboratory, University of Bristol, Tyndall Avenue, Bristol, BS8 1TL, United Kingdom}

\author{C. Putzke}
\altaffiliation{Present address: Laboratory of Quantum Materials, Institute of Materials, \'Ecole Polytechnique F\'ed\'erale de Lausanne (EPFL), 1015 Lausanne, Switzerland}
\affiliation{H. H. Wills Physics Laboratory, University of Bristol, Tyndall Avenue, Bristol, BS8 1TL, United Kingdom}

\author{D. Kaczorowski}
\affiliation{Institute of Low Temperature and Structure Research, Polish Academy of Sciences, 50-950 Wroclaw, Poland}

\author{T. Wolf}
\affiliation{Institute for Quantum Materials and Technologies, Karlsruhe Institute of Technology, 76021 Karlsruhe, Germany}

\author{F. Hardy}
\affiliation{Institute for Quantum Materials and Technologies, Karlsruhe Institute of Technology, 76021 Karlsruhe, Germany}

\author{C. Meingast}
\affiliation{Institute for Quantum Materials and Technologies, Karlsruhe Institute of Technology, 76021 Karlsruhe, Germany}

\author{J. G. Analytis}
\altaffiliation{Present address: Department of Physics, University of California, Berkeley, California 94720, USA.}
\affiliation{Geballe Laboratory for Advanced Materials and Department of Applied Physics, Stanford University, California 94305-4045, USA}
\affiliation{Stanford Institute for Materials and Energy Sciences, SLAC National Accelerator Laboratory, 2575 Sand Hill Road, Menlo Park, California 94025, USA}

\author{J.-H. Chu}
\altaffiliation{Present address: Department of Physics, University of Washington, Seattle, WA, 98195 USA}
\affiliation{Geballe Laboratory for Advanced Materials and Department of Applied Physics, Stanford University, California 94305-4045, USA}
\affiliation{Stanford Institute for Materials and Energy Sciences, SLAC National Accelerator Laboratory, 2575 Sand Hill Road, Menlo Park, California 94025, USA}

\author{I.R. Fisher}
\affiliation{Geballe Laboratory for Advanced Materials and Department of Applied Physics, Stanford University, California 94305-4045, USA}
\affiliation{Stanford Institute for Materials and Energy Sciences, SLAC National Accelerator Laboratory, 2575 Sand Hill Road, Menlo Park, California 94025, USA}

\author{A. Carrington}
\email{Corresponding author: a.carrington@bristol.ac.uk}
\affiliation{H. H. Wills Physics Laboratory, University of Bristol, Tyndall Avenue, Bristol, BS8 1TL, United Kingdom}

\begin{abstract}
A long-standing theoretical prediction is that in clean, nodal unconventional superconductors the magnetic penetration depth $\lambda$, at zero temperature, varies linearly with magnetic field.  This non-linear Meissner effect is an equally important manifestation of the nodal state as the well studied linear-in-$T$ dependence of $\lambda$, but has never been convincingly experimentally observed. Here we present measurements of the nodal superconductors CeCoIn$_5$ and LaFePO which clearly show this non-linear Meissner effect.  We further show how the effect of a small dc magnetic field on $\lambda(T)$ can be used to distinguish gap nodes from non-nodal deep gap minima.  Our measurements of KFe$_2$As$_2$ suggest that this material has such a non-nodal state.
\end{abstract}
\maketitle

\section{Introduction}

Determination of the symmetry and momentum dependent structure of the superconducting energy gap $\Delta(\bm{k})$ is of fundamental importance as this provides a strong guide to microscopic theories of superconductivity \cite{Hirschfeld2011}. Measurements of the temperature dependence of the magnetic penetration depth $\lambda(T)$ have proved extremely useful as $\lambda(T)$ is directly related to the energy dependence of the quasiparticle density of states $N(E)$ which in turn is related to $\Delta(\bm{k})$ \cite{Prozorov2006}. For example, a line node in $\Delta(\bm{k})$ on a quasi-two-dimensional Fermi surface causes $N(E)$ to vary linearly with energy and $\lambda$ to vary linearly with $T$. In contrast, if $\Delta(\bm{k})$ is finite for all $\bm{k}$ then $\lambda(T)$ will vary exponentially for $T$ much less than the minimum gap \cite{Carrington2011}. 

It was theoretically proposed \cite{Yip1992} that the magnetic field dependence of $\lambda$ also depends strongly on $\Delta(\bm{k})$ and thus provides an alternative test of the pairing state.  In a nodal superconductor in the clean limit, the theory for this non-linear Meissner effect predicts that $\lambda$ is a linear function of $H$ at $T=0$, with a field scale $H_0$ which is of order the thermodynamic critical field $H_c$ (i.e., $\Delta \lambda (H)/\lambda_0 = H/H_0$).  However, despite considerable experimental effort, this long-standing prediction has not been convincingly observed experimentally. Although the  change in $\lambda$ with field, $\Delta \lambda(H)$, was found to be approximately linear in the cuprate superconductor YBa$_2$Cu$_3$O$_{6+x}$ (Y123) \cite{Carrington1999,Bidinosti1999}, $\Delta \lambda(H)$ had a very weak temperature and angle dependence. The latter might be explained by the orthorhombicity of Y123 \cite{Halterman2001} but the lack of temperature dependence of $\Delta\lambda(H)$ is in serious disagreement with theory and strongly suggests that the observed effects in Y123 were of extrinsic origin \cite{Carrington1999,Bidinosti1999}. Attempts to observe the effect in Y123 through transverse torque measurements  were also unsuccessful \cite{Bhattacharya1999}.  Later it was predicted \cite{Dahm97} that the non-linear current relation which leads to the linear $\lambda(H)$ in nodal superconductors should also give rise to intermodulation distortion (IMD) in the microwave response.  Although extrinsic defects also give rise to IMD, the theoretically predicted increase in amplitude of the IMD as temperature is decreased was observed in high quality films of Y123 \cite{Oates2004} suggesting that the intrinsic response is present and adds to the extrinsic IMD.  Despite the IMD results, the lack of experimental evidence for the predicted response of $\lambda$ with field remains an outstanding problem which could suggest a problem with the standard quasiclassical theory of unconventional superconductivity.

Here we present measurements of $\lambda(H)$ for two putative nodal superconductors; CeCoIn$_5$ and LaFePO which show convincingly that the predicted field dependence of $\lambda$ is present in nodal superconductors and this field dependence decreases with increasing temperature as expected.  Our measurements of a third material,  KFe$_2$As, are markedly  different which we argue results from a small but finite non-nodal gap in this superconductor.  We show how the combination of the effect of field and temperature on $\lambda$ can be used to distinguish between the case where there is a small density of impurities in a nodal superconductor from the case where there is a small but finite energy gap.

CeCoIn$_5$ has been extensively studied using many probes including specific heat, thermal conductivity $\kappa(T)$ \cite{Movshovich2001,Seyfarth2008} and $\lambda(T)$ \cite{Ormeno2002,Chia2003,Ozcan2003} with the results consistent with $d$-wave superconductivity with line nodes. For LaFePO, $\lambda(T)$ \cite{Fletcher2009,Hicks2009} and $\kappa(T)$ \cite{Yamashita2009,Sutherland2012} measurements indicate line nodes but with theory suggesting $\Delta(\bm{k})$ to have $A_{2g}$ ($s$-wave) symmetry \cite{Kuroki2009}. KFe$_2$As$_2$ is more controversial; although $\lambda(T)$ has a strong, quasi-linear dependence consistent with gap nodes \cite{Hashimoto2010}, specific heat measurements \cite{Hardy2013} show evidence for very small gaps which may be difficult to distinguish from true nodes. $\kappa(T)$ measurements have been argued to show universal behaviour consistent with $d$-wave order \cite{Reid2012}, although this has been disputed \cite{Watanabe2014}. Additionally, angle-resolved photoemission (ARPES) measurements have indicated the presence of eight-fold nodes on one of the Fermi surface sheets \cite{Okazaki2012}. This result should be viewed with some caution since the measurement was carried out at $T = 1.5$ K, and given $T_c = 3.4$ K, the gap structure will not have fully developed to its zero temperature value. Studies of $\lambda(T)$ and the effect of electron irradiation on the doped series Ba$_{1-x}$K$_x$Fe$_2$As$_2$, in combination with theoretical work, suggest that Ba$_{1-x}$K$_x$Fe$_2$As$_2$ has a highly anisotropic gap which may change sign on some parts of the Fermi surface as $x\rightarrow 1$ \cite{Cho2016}. Hence it remains an open question whether KFe$_2$As$_2$ has gap nodes or not.

\begin{figure}
\centering
\includegraphics[width=0.8\linewidth]{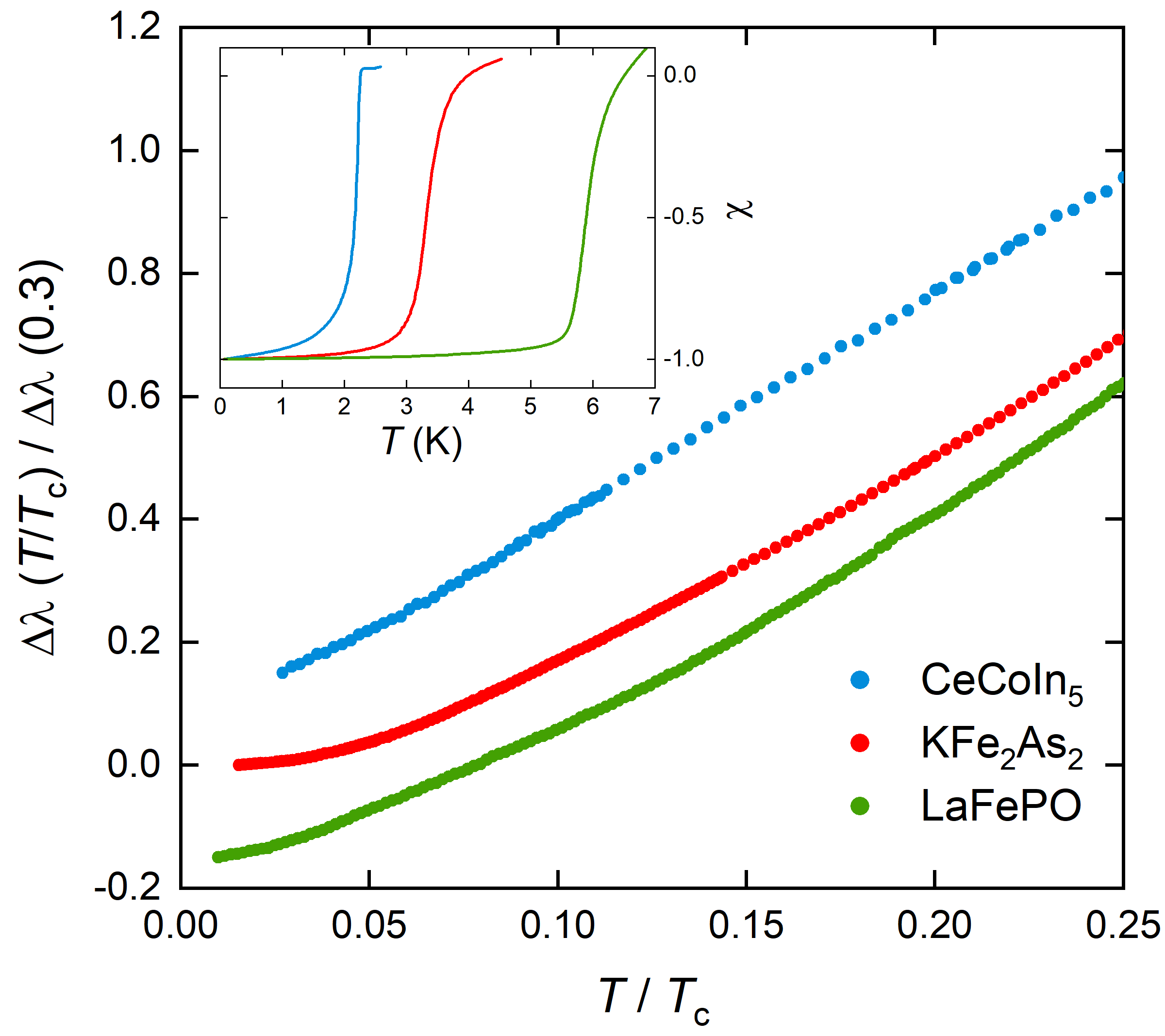}
\caption{\textbf{Temperature dependence of in-plane $\lambda$ relative to its value at the lowest temperature $\Delta\lambda(T)$ for CeCoIn$_5$, LaFePO and KFe$_2$As$_2$}. For each material $\Delta\lambda(T)$ is normalized by its value at $T/T_c=0.3$. The data for CeCoIn$_5$, and LaFePO have been shifted by $\pm$0.15 along the $\lambda$ axis for clarity. The inset shows the RF susceptibility over the full temperature range to emphasise the superconducting transitions. Here $\chi$ was set to zero just above $T_c$ and normalized to $-1$ at the lowest temperature. The mid-points of the transition are 2.1\,K, 3.2\,K and 5.9\,K for CeCoIn$_5$, KFe$_2$As$_2$ and LaFePO respectively. For KFe$_2$As$_2$ and CeCoIn$_5$ the RF field is parallel to the $ab$ plane, and for LaFePO the RF field is parallel to the $c$ axis.} 
\label{Fig:LambdaT}
\end{figure}

\section{Results}

\textbf{Temperature dependence of the zero field penetration depth}. The temperature dependence of the in-plane $\lambda$ for samples of  CeCoIn$_5$, LaFePO and  KFe$_2$As, in zero applied dc field and negligibly small ac probe field, is shown in Fig.\ \ref{Fig:LambdaT}. All three materials show a predominately linear temperature dependence to $\lambda$ for $T \ll T_c$ but below some lower temperature $T^*$ the temperature dependence flattens off considerably (less so in the case of CeCoIn$_5$). The behaviour of all three compounds is consistent with deep gap minima, which may or may not be nodal, in the presence of a finite density of impurities.

For CeCoIn$_5$ and KFe$_2$As$_2$ the measurements were performed with $H$ parallel to the $ab$-plane of the material to minimize demagnetizing effects. As the samples are thin and neither material is very strongly anisotropic the measured $\Delta\lambda(T)$ in this geometry is predominately the in-plane response but will also contain a small contribution from the out-of-plane $\lambda(T)$ (see SI). For LaFePO, we used the $H\|c$ geometry because of the larger anisotropy of $\lambda(T)$ and the thicker $c$-axis dimension of the sample.

The results for KFe$_2$As$_2$ and LaFePO are similar to those reported previously \cite{Hashimoto2010,Fletcher2009} but here are extended to lower temperature. For CeCoIn$_5$, the response is linear-in-$T$ down to lower $T$ than the other two compounds, suggesting a lower concentration of impurities. Previous measurements of $\lambda(T)$ in CeCoIn$_5$ \cite{Ormeno2002,Chia2003,Ozcan2003} and particularly Ref.\ \cite{Hashimoto2013} have shown $\lambda(T)$ following a power law ($T^n$) with exponent closer to $1.5$, which was interpreted in Ref.\ \cite{Hashimoto2013} as evidence of proximity to a quantum critical point.  However, the present results suggest that the higher value of $n$ may have been caused by surface impurities which we removed by cleaving (see method).  A $T^{1.5}$ power law is difficult to distinguish from the response of an impure $d$-wave superconductor \cite{Carrington1999}.

\begin{figure}
\centering
\includegraphics[width=0.5\linewidth]{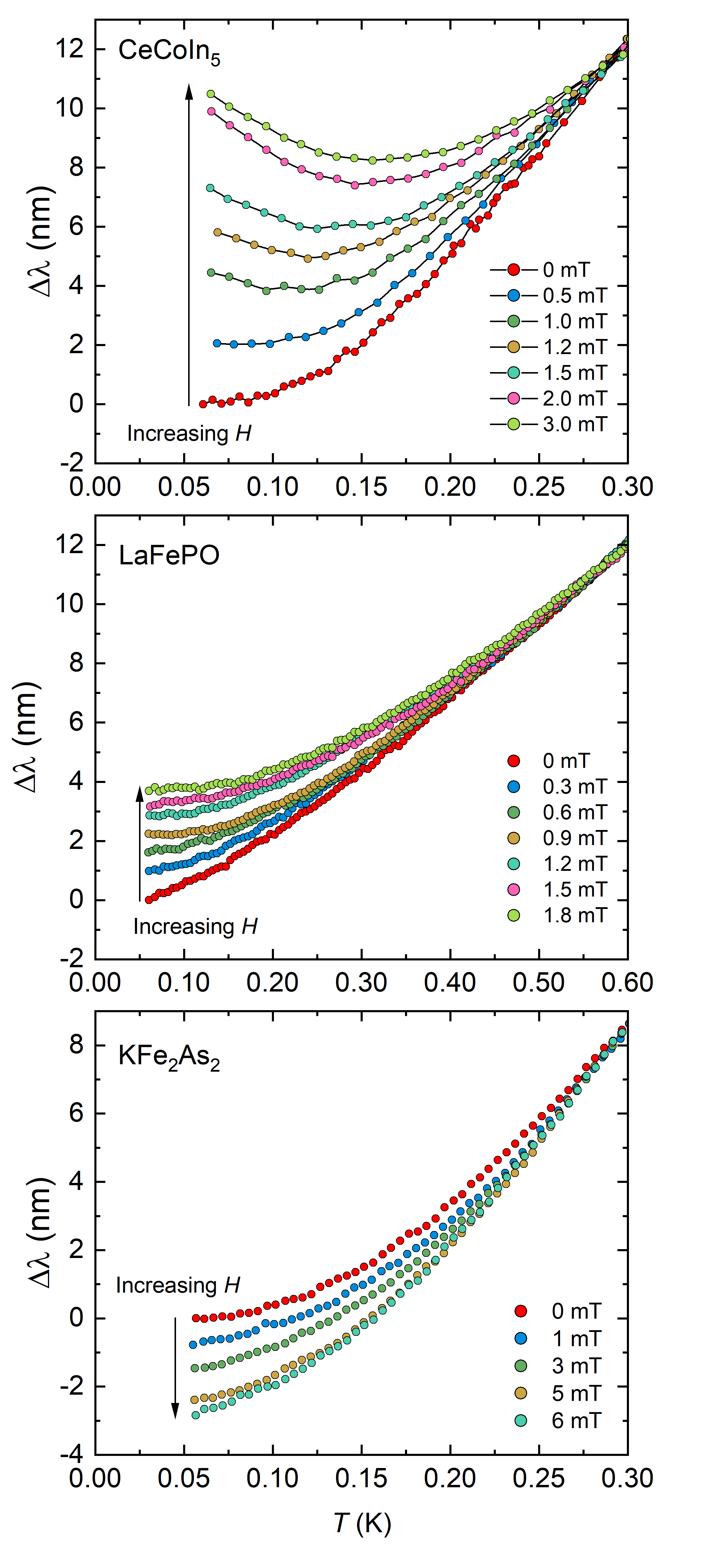}
\caption{\textbf{Effect of magnetic field on the temperature dependence of $\Delta\lambda$}. The finite field data have been shifted along the $\lambda$ axis to coincide with the zero field result at $T=0.3$\,K for CeCoIn$_5$ and KFe$_2$As$_2$ and $T=0.6$\,K for LaFePO, in order to emphasise the progressive change in the temperature dependence of $\lambda$ with field as in Fig.\ \ref{Fig:calc}.}
\label{Fig:LambdaTB}
\end{figure}

\textbf{Magnetic field dependence of the penetration depth}. In Fig.\ \ref{Fig:LambdaTB} we show the effect of a small dc field on $\lambda(T)$ for all three materials.  Here we have normalised the results in finite field to coincide with the zero field data at the highest temperature in the figure which highlights the systematic changes in $\lambda(T)$ with field, and removes any field dependent background effects (see methods).  For both CeCoIn$_5$ and LaFePO, $\lambda(T)$ becomes less temperature dependent with increasing field so that the normalised $\Delta\lambda(T)$ increases with field at the lowest temperature.   For KFe$_2$As$_2$, the effect of $H$ is opposite with the temperature dependence of $\lambda(T)$ becoming stronger so that $\Delta\lambda$ decreases with $H$ at the lowest temperature.  The materials remain in the Meissner state for all $T$ shown in this figure as  $H<H_p(T)$ (the field of first flux penetration which was measured by a Hall probe magnetometer or SQUID magnetometer for each sample, see SI).   For LaFePO, measured in the $H\|c$ geometry, the surface fields will be non-uniform because of demagnetising effects, however the effect of this on the measured surface averaged $\lambda(H)$ is estimated to be minimal (see SI).  

\begin{figure}
\centering
\includegraphics[width=0.8\linewidth]{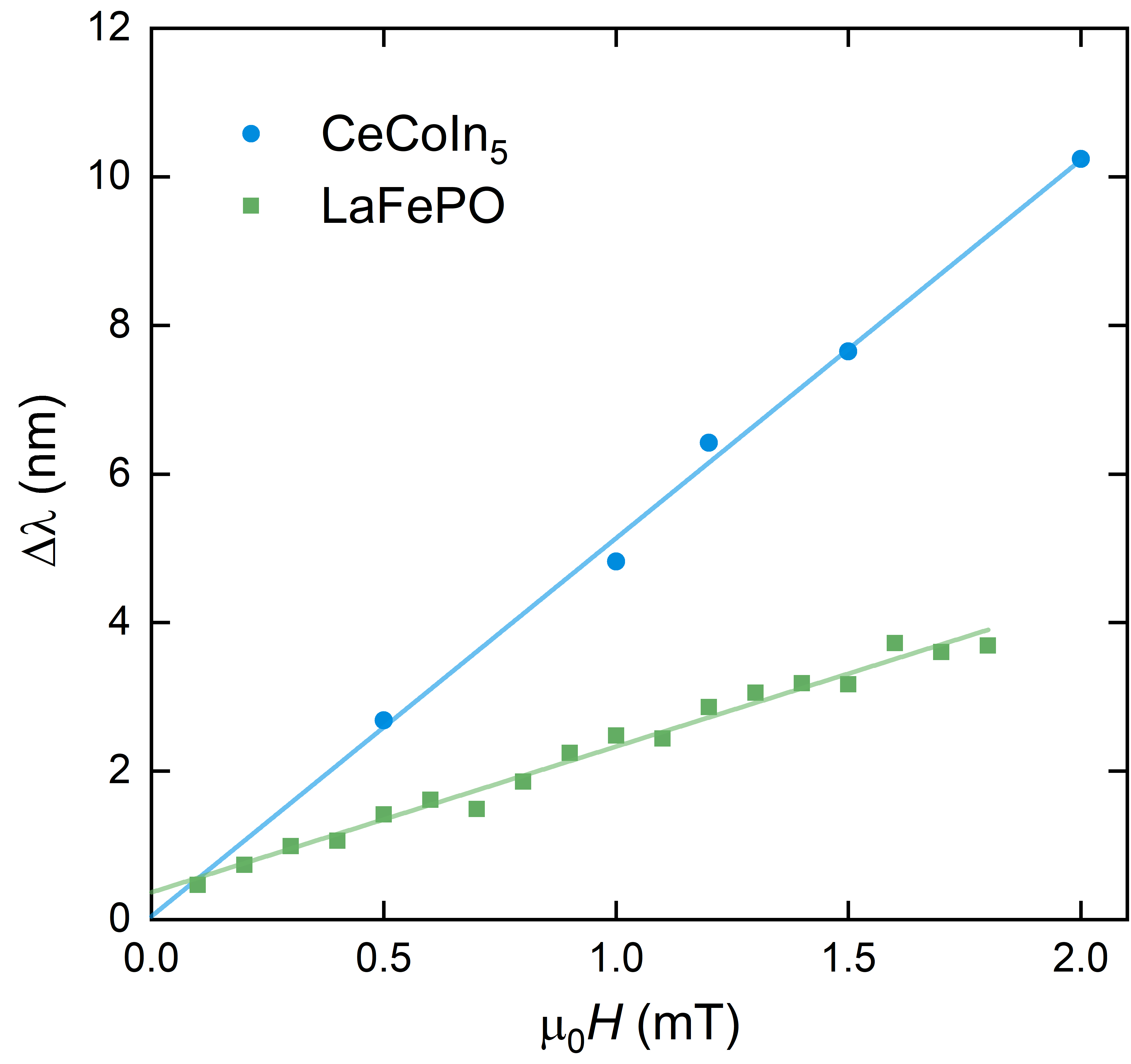}
\caption{\textbf{Field dependence of $\Delta\lambda$ at the lowest temperatures} - (65\,mk, 65\,mk and 55\,mk for CeCoIn$_5$, LaFePO and KFe$_2$As$_2$ respectively, relative to the change at $T$=0.3\,K (0.6\,K for LaFePO)). The lines are linear fits to the data.}
\label{Fig:LambdaB}
\end{figure}

In Fig.\ \ref{Fig:LambdaB} we show that for both CeCoIn$_5$ and LaFePO the change in $\lambda$ with $H$ at the base temperature relative to the change at our reference temperature varies linearly with $H$.  As $\Delta\lambda(H)$ decreases strongly with temperature (Fig.\ \ref{Fig:LambdaTB}), normalising at this higher reference temperature should have only a small impact on $\Delta\lambda(H)$ at the lowest temperatures. The observed linear-in-$H$ behaviour of $\Delta\lambda(H)$ is in excellent agreement with theory for the response of a clean, nodal superconductor - thus confirming the key prediction of Ref.\ \cite{Yip1992}. A direct comparison with the theory can be made by comparing the temperature and field dependent results in Fig.\ \ref{Fig:LambdaTB} to theoretical calculations in Fig.\ \ref{Fig:calc} where the results have been normalised in the same way.  The magnitude of $\Delta\lambda(H)$ is in broad agreement with estimates (see SI) using the material parameters, although a more quantitative analysis requires more detailed theoretical modelling including multiband effects.  

\section{Discussion}

\begin{figure}
\includegraphics[width=0.9\linewidth]{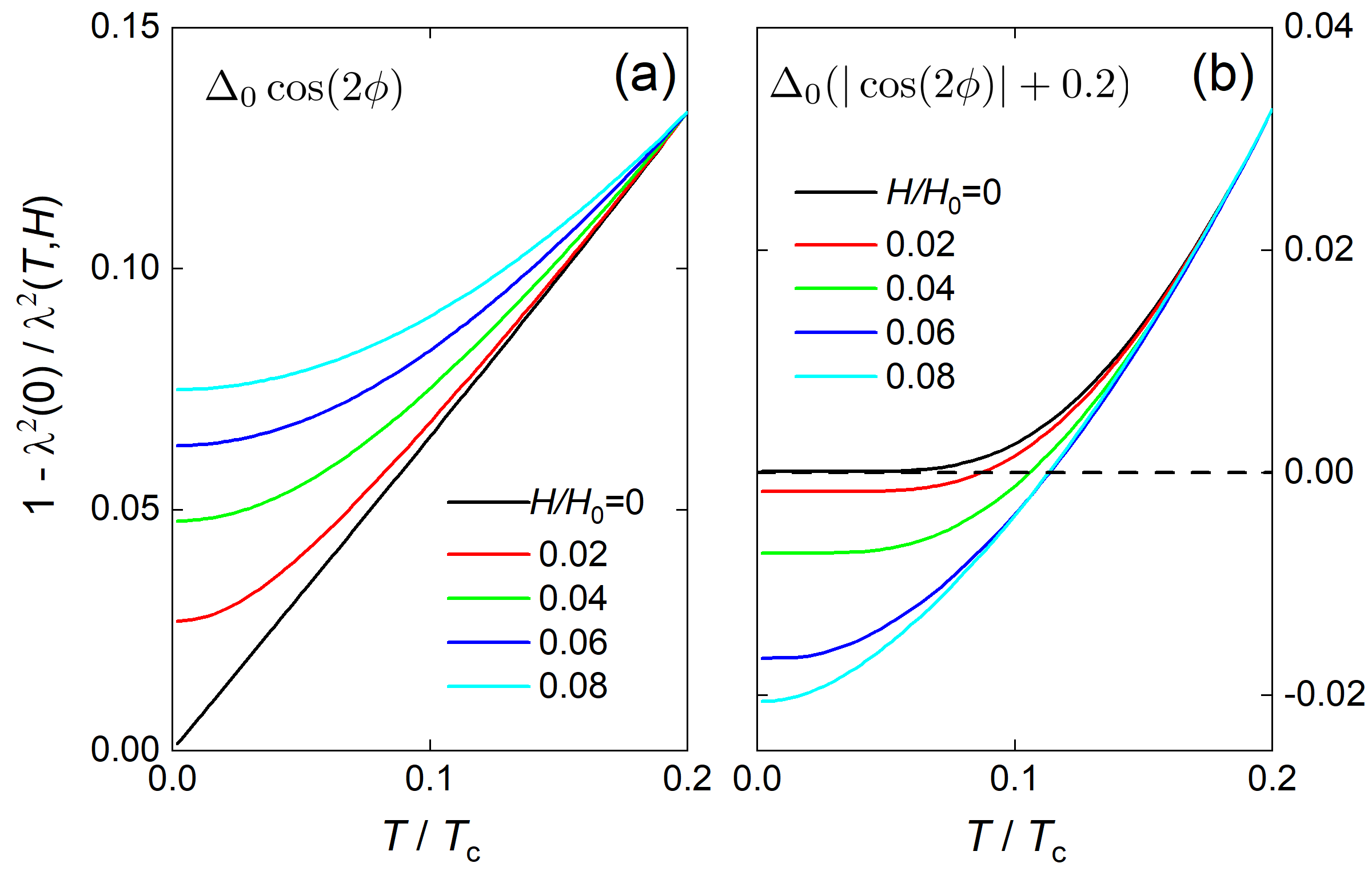}
\caption{\textbf{Calculated temperature dependence of the normalized superfluid density $\lambda^2(0)/\lambda^2(T,H)$}. (a) $d$-wave gap structure (b) Gap structure is strongly anisotropic but with small finite gap. Both cases are in the clean-limit and the finite field results have been shifted vertically so that they coincide with the $H=0$ results at $T/T_c=0.2$ for comparison with our experimental results. Unshifted results and details of the calculations are given in the SI. Note: $1-\lambda^2_0/\lambda^2(T,H) \simeq 2\Delta\lambda(T,H)/\lambda_0$ for small $\Delta\lambda(T,H)$.}
\label{Fig:calc}
\end{figure}

A problem with interpreting zero field $\lambda(T)$ to determine $\Delta(\bm{k})$ is the effect of impurities. Non-magnetic impurities produce gapless excitations for a range of $\bm{k}$ around a node and $\lambda(T) \sim T^2$ below a characteristic temperature, $T^*$ which depends on the impurity density \cite{hirschfeld1993}. Paramagnetic impurities also cause complications as these can cause upturns or flattening of $\lambda(T)$ at low temperature \cite{Cooper1996}. In practice, it can be difficult to distinguish the case where there is a small density of impurities in a nodal superconductor from the case where there is a small but finite energy gap and so the results in Fig.\ \ref{Fig:LambdaT} on their own do not provide decisive evidence for a nodal or non-nodal gap.  This distinction is important, because a node is strong evidence that $\Delta(\bm{k})$ changes sign on one or more Fermi surface sheet, which can prove to be strong evidence used to differentiate between different microscopic models of superconductivity. 

The field dependence of $\lambda(T)$ provides a route to distinguish between the nodal and non-nodal cases which is less sensitive to impurities than the zero field temperature dependence of $\lambda$ alone.  The non-linear Meissner effect arises from the Doppler shift of the quasiparticle energies in a field, $\delta \varepsilon \propto \bm{v_s\cdot v_F}$ where $\bm{v_s}$ is the superflow velocity of the screening currents and $\bm{v_F}$ is the Fermi velocity. Close to a node in $\Delta(\bm{k})$ this shifts the quasiparticle states below the Fermi level and hence they become occupied, producing backflow jets which reduce the effective superfluid density.  At finite temperature, the field dependence of $\lambda$ is reduced for $H < H^*$, where $H^*\sim  H_0T/T_c$, so that at fixed $H$, the temperature dependence of $\lambda$ becomes weaker at low $T$ (see Fig.\ \Ref{Fig:calc}(a)). In the case where the gap is isotropic, at low temperatures and small fields the shift in energy of the quasiparticle states is small compared to the gap for all $\bm{k}$ and so the effect is essentially absent \cite{Xu1995}.  However, when the gap is anisotropic but non-nodal, then close to the momentum where $\Delta(\bm{k})$ is minimum, this shift will move the minimum closer to zero, making the gap smaller, and hence at sufficiently low temperature and for weak fields $\Delta \lambda(T)$ will have a \textit{stronger} $T$-dependence compared to zero field (see Fig.\ \ref{Fig:calc} (b)).  This distinct, contrasting change in the temperature dependence of $\lambda$ with field is a key distinguishing factor between the nodal and non-nodal gap structures. Importantly, this difference is robust in the presence of a low density of impurities, both magnetic and non-magnetic (see below).

The markedly different behavior of KFe$_2$As$_2$ compared to the other two materials, shown in Fig.\ \ref{Fig:LambdaTB}, can be understood if $\Delta\lambda(H)$ is dominated by a small but finite energy gap, such as that modelled in Fig.\ \ref{Fig:calc}(b).  In a multiband material like KFe$_2$As$_2$, there may be both nodal and small-finite gapped sheets of Fermi surface \cite{Hardy2013}, and the measured response would be the sum of the superfluid density from all sheets. It is possible, therefore, that there could be a small contribution from a nodal sheet with sufficient impurity scattering to reduce its contribution to $\lambda(H)$. However, our results show that the non-linear response is dominated by one or more sheets with a small but finite gap, which rules out there being nodes on all sheets.  Further modelling could consider how impurities, causing inter- and intra-band scattering, would change the gap structure and hence the non-linear response.

\begin{figure}
\centering
\includegraphics[width=0.8\linewidth]{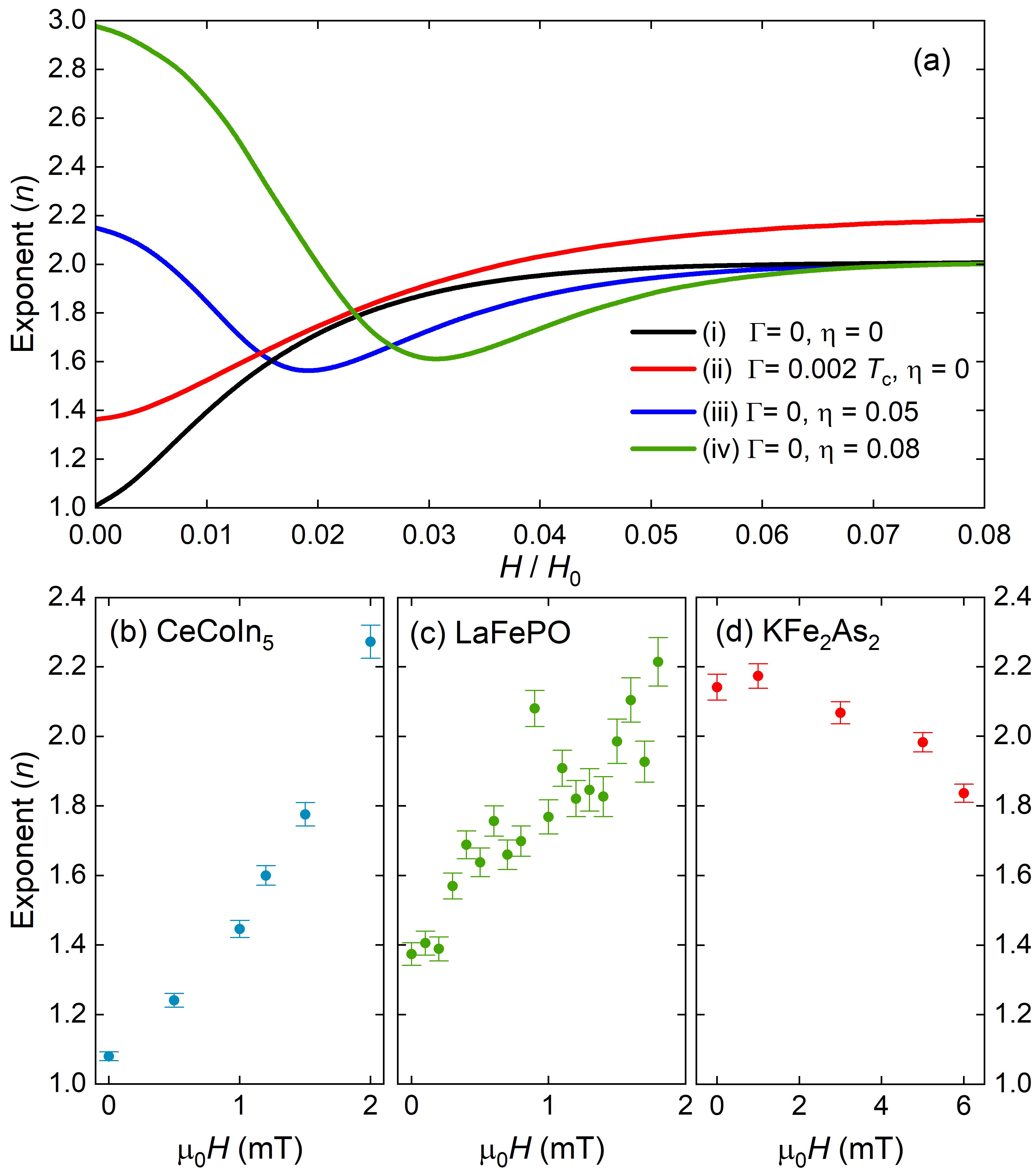}
\caption{\textbf{Power law exponent analysis for theory and experimental data}. (a) Evolution of the exponent with field from fits to the theoretical response for different gap structures of the form $\Delta(\phi,T)=\Delta_0(T) (|\cos(2 \phi)| + \eta)$ and varying impurity concentration $\Gamma$ (see SI for details of the calculations). (b - d) Experimental response from CeCoIn$_5$, LaFePO and KFe$_2$As$_2$ respectively. The exponent $n$ is extracted from fits to $\lambda^2_0/[\lambda_0+\Delta\lambda(T)]^2=1-AT^n$ with upper $T$ limit of  300\,mK for CeCoIn$_5$ and KFe$_2$As$_2$, and up to 400\,mK for LaFePO.  The values assumed for $\lambda_0$ were 190\,nm for CeCoIn$_5$ \cite{Uemura2009}, 240\,nm for LaFePO \cite{Ormeno2002} and 200\,nm for KFe$_2$As$_2$ \cite{Furukawa2011}. For the fits to the theoretical response an upper limit of $T/T_c=0.1$ was used.}
\label{fig:exponent}
\end{figure}

An alternative way to analyse the data is to focus directly on the temperature dependence of $\lambda$ rather than its field dependent changes at a fixed temperature relative to that at some higher $T$ reference point (as we have done in Figs\ \ref{Fig:LambdaTB} and \ref{Fig:LambdaB}).   In order to do this we fit a power-law to the normalised superfluid density; $\tilde{\rho}\equiv \lambda^2_0/[\lambda_0+\Delta\lambda(T)]^2=1-AT^n$ at low temperature.  We fit  $\tilde{\rho}$ rather than $\Delta\lambda$ because in the nodal case the former remains closer to the asymptotic low temperature behaviour over a much larger range of temperature \cite{Carrington1999}. The resulting exponent $n$ is only weakly dependent on the assumed value of $\lambda_0$ \cite{Fletcher2009}. The fits were performed over the temperature range from base temperature up to 300\,mK for CeCoIn$_5$ and KFe$_2$As$_2$, and up to 400\,mK for LaFePO. Note that this power-law behaviour is only an approximation to the true exponential behaviour at low temperature and zero field for the case of a finite-gap, but is frequently used to fit experimental data \cite{Cho2016}. The extracted exponent also depends on the temperature range of the fit even for the nodal case, as a true power-law is theoretically only found in the asymptotic low-$T$ behaviour. For KFe$_2$As$_2$, $n(H=0)$ appears to be higher than what is reported in Ref.\ \cite{Cho2016} ($n \simeq 1.45$), however, a direct comparison between the two sets of data shows excellent agreement (see SI). The discrepancy in the value of $n$ simply results from the different $T$ ranges of the fits. 

Fig.\ \ref{fig:exponent}(b-d) shows how the exponent $n$ varies with $H$ for our three compounds.  At zero field, CeCoIn$_5$ has $n$ close to 1, consistent with line nodes in the clean limit. As $H$ increases, $n$ increases too and slightly exceeds 2 at the highest field.  For LaFePO the behaviour is similar, but with $n$ higher than 1 in zero field, consistent with a small amount of impurities.  Once again, KFe$_2$As$_2$ shows contrasting behaviour with $n$ being close to 2 in zero field and $n$ decreasing with increasing $H$.  Hence, for CeCoIn$_5$ and LaFePO $\tilde{\rho}$ becomes less temperature dependent at low $T$ in increasing field, whereas for KFe$_2$As$_2$ the opposite is true.

This experimental behaviour can be compared directly with a similar analysis of our calculations of the non-linear response.  In Fig.\ \ref{fig:exponent}(a) we show the evolution of $n$ for a nodal gap structure with and without impurities, together with the response for finite-gap structure (with two different values of finite gap).  For the pure $d$-wave case (\textit{i}) , $n$ increases from 1 to 2 as $H$ is increased, with the majority of the increase occurring for $H\ll H_0$ ($n=1.8$ for  $H/H_0=0.025$). The addition of impurities (case (\textit{ii})) increases $n(H=0)$ and also causes $n$ to more slowly approach its high field limit which is now slightly above 2.  Cases (\textit{iii}) and (\textit{iv}) simulate two different sizes of finite gap. In both cases,  $n$ begins above 2 and then decreases, reaching a minimum before increasing and finally saturating at 2 at high field.

The experimental data for CeCoIn$_5$ is close to the clean nodal-wave response [case (\textit{i})] and for LaFePO it is similar to impure nodal response [case (\textit{ii})].  For KFe$_2$As$_2$ the decrease in $n$ is again shown to be similar to the finite gap cases (\textit{iii}) and (\textit{iv}) although it was not possible to extend the field to high enough values to observe the predicted minimum in $n$ because this is limited by $H_p$ or $H_{c1}$.

Our measurements of $\lambda(H,T)$ in CeCoIn$_5$ and LaFePO provide the first unambiguous observation of the theoretically predicted \cite{Yip1992,Stojkovic1995,Xu1995} field effect on the magnetic penetration depth in nodal superconductors. At the lowest temperatures $\Delta\lambda(H)$ is linear-in-$H$ and its magnitude decreases strongly with increasing temperature as predicted.  Although the effect may be too small to be clearly seen in the cuprates, the response in lower $T_c$ unconventional superconductors is much stronger. This is because the maximum size of $\lambda(H)$ is determined by the ratio $H_{c1}/H_0$ and is approximately proportional to $T_c^{-1}$ and hence the response is much larger for low-$T_c$ materials than for optimally doped cuprates. Furthermore, the bulk non-linear response in cuprates can be masked by zero-energy surface Andreev bound states  \cite{Carrington2001, Barash2000, Zare2010}, which should not be present (at zero energy) in multiband materials with sign changing gap such as iron-based superconductors \cite{Araujol2009} or CeCoIn$_5$. 

We have shown how the field dependence of the magnetic penetration depth can be used to distinguish between nodal and small finite gap structures. Magnetic field causes $\lambda(T)$ in the former case to become less temperature dependent at low temperature ($T\ll T_c$), whereas an increased temperature dependence is found for the latter case. The results confirm that both CeCoIn$_5$ and LaFePO have a sign changing nodal gap structure, whereas for KFe$_2$As$_2$ the response is dominated by a small but finite gap, restricting possible gap symmetries.  With further modelling of the non-linear response for multiband and multigap superconductors in the presence of impurities it should be possible to extract more quantitative information on the size of the minimum energy gaps and the limits to any possible contribution from sheets with gap nodes. As measurements of the field dependence of $\lambda$ are relatively simple to perform by adding a dc field to the widely-used tunnel-diode penetration depth apparatus, we expect that it will prove to be a very useful addition to the tool-kit used to investigate gap symmetry in other candidate nodal superconductors.

\section{Methods}

\textbf{Sample growth and characterisation}. Samples of KFe$_2$As$_2$ were grown by a self flux technique and are similar to those used in a previous heat capacity study \cite{Hardy2013}. We found that the measured $\lambda(T)$ was very sensitive to the surface quality and exposing them to air even for a short time changed $\lambda(T)$ markedly, producing a fully gapped response \cite{Wilcox2019}. Hence, the samples were cleaved on all six sides in an argon glove box and covered with degassed grease at all times when outside of this. The CeCoIn$_5$ samples were grown in an In flux.  Long term exposure to air resulted in visible surface corrosion and so the samples were also cleaved and covered with grease prior to measurement. The LaFePO samples were grown in a Sn flux and are not air sensitive so were not cleaved.

\textbf{Measurements of the magnetic penetration depth}. Our measurements of $\lambda(T,H)$ were performed using a radio frequency (RF) tunnel diode oscillator technique \cite{Carrington1999}. The sample is attached to a sapphire cold finger and inserted into a small, copper solenoid which forms part of a tank circuit that is resonated at 14\,MHz using the tunnel diode circuit. The sapphire cold finger is cooled down to a minimum temperature of 50\,mK using a dilution refrigerator. The RF field is estimated to be a few $\mu T$ and the Earth's field is shielded using a mu-metal can.

For the field dependent measurements, a small dc field is applied parallel to the RF field using a superconducting solenoid. The samples are first cooled to base temperature ($\sim 50$\,mK) in zero field before applying the dc field in order to minimize any effect of vortices entering the sample and contributing to $\lambda(T)$. The samples were then warmed to 0.3\,K (0.6\,K for LaFePO) and cooled back to base temperature several times to average the data and remove any thermal drift.

In each case, we set the zero for $\Delta \lambda(T)$ to be the lowest temperature ($T_{\mathrm{min}} \sim 50$\,mK) for the zero field measurement, \emph{i.e.}\ $\Delta \lambda (T_{\mathrm{min}}, H = 0) = 0$. The measurements in finite field are then shifted in $\lambda$ so that they coincide with the zero field measurements at $T=0.3$\,K (or $T=0.6$\,K for LaFePO). This process is necessary because applying a dc field causes shifts in frequency of the resonant circuit which are not due to the sample and are difficult to subtract reliably.  By fixing the dc field and sweeping the sample temperature only, the response of the coil is not measured.

\section{Acknowledgements}

This work was supported by EPSRC grants EP/R011141/1, EP/L015544/1, and EP/H025855/1. Crystal growth and characterization of LaFePO at Stanford University was supported by the Department of Energy, Office of Basic Energy Sciences under Contract No. DE-AC02-76SF00515. Crystal growth and characterization of CeCoIn$_5$ was supported by the National Science Centre (Poland) under research grant No. 2015/19/B/ST3/03158.

%\bibliographystyle{apsrev}
%\bibliography{../NLME2019}

%apsrev4-2.bst 2019-01-14 (MD) hand-edited version of apsrev4-1.bst
%Control: key (0)
%Control: author (8) initials jnrlst
%Control: editor formatted (1) identically to author
%Control: production of article title (0) allowed
%Control: page (0) single
%Control: year (1) truncated
%Control: production of eprint (0) enabled
\begin{thebibliography}{0}%
\makeatletter
\providecommand \@ifxundefined [1]{%
 \@ifx{#1\undefined}
}%
\providecommand \@ifnum [1]{%
 \ifnum #1\expandafter \@firstoftwo
 \else \expandafter \@secondoftwo
 \fi
}%
\providecommand \@ifx [1]{%
 \ifx #1\expandafter \@firstoftwo
 \else \expandafter \@secondoftwo
 \fi
}%
\providecommand \natexlab [1]{#1}%
\providecommand \enquote  [1]{``#1''}%
\providecommand \bibnamefont  [1]{#1}%
\providecommand \bibfnamefont [1]{#1}%
\providecommand \citenamefont [1]{#1}%
\providecommand \href@noop [0]{\@secondoftwo}%
\providecommand \href [0]{\begingroup \@sanitize@url \@href}%
\providecommand \@href[1]{\@@startlink{#1}\@@href}%
\providecommand \@@href[1]{\endgroup#1\@@endlink}%
\providecommand \@sanitize@url [0]{\catcode `\\12\catcode `\$12\catcode
  `\&12\catcode `\#12\catcode `\^12\catcode `\_12\catcode `\%12\relax}%
\providecommand \@@startlink[1]{}%
\providecommand \@@endlink[0]{}%
\providecommand \url  [0]{\begingroup\@sanitize@url \@url }%
\providecommand \@url [1]{\endgroup\@href {#1}{\urlprefix }}%
\providecommand \urlprefix  [0]{URL }%
\providecommand \Eprint [0]{\href }%
\providecommand \doibase [0]{https://doi.org/}%
\providecommand \selectlanguage [0]{\@gobble}%
\providecommand \bibinfo  [0]{\@secondoftwo}%
\providecommand \bibfield  [0]{\@secondoftwo}%
\providecommand \translation [1]{[#1]}%
\providecommand \BibitemOpen [0]{}%
\providecommand \bibitemStop [0]{}%
\providecommand \bibitemNoStop [0]{.\EOS\space}%
\providecommand \EOS [0]{\spacefactor3000\relax}%
\providecommand \BibitemShut  [1]{\csname bibitem#1\endcsname}%
\let\auto@bib@innerbib\@empty
%</preamble>
\end{thebibliography}%


\begin{thebibliography}{26}
\expandafter\ifx\csname natexlab\endcsname\relax\def\natexlab#1{#1}\fi
\expandafter\ifx\csname bibnamefont\endcsname\relax
  \def\bibnamefont#1{#1}\fi
\expandafter\ifx\csname bibfnamefont\endcsname\relax
  \def\bibfnamefont#1{#1}\fi
\expandafter\ifx\csname citenamefont\endcsname\relax
  \def\citenamefont#1{#1}\fi
\expandafter\ifx\csname url\endcsname\relax
  \def\url#1{\texttt{#1}}\fi
\expandafter\ifx\csname urlprefix\endcsname\relax\def\urlprefix{URL }\fi
\providecommand{\bibinfo}[2]{#2}
\providecommand{\eprint}[2][]{\url{#2}}

\bibitem[{\citenamefont{Howald et~al.}(2013)\citenamefont{Howald, Maisuradze,
  de~R{\'e}otier, Yaouanc, Baines, Lapertot, Mony, Brison, and
  Keller}}]{Howald2013}
\bibinfo{author}{\bibfnamefont{L.}~\bibnamefont{Howald}},
  \bibinfo{author}{\bibfnamefont{A.}~\bibnamefont{Maisuradze}},
  \bibinfo{author}{\bibfnamefont{P.~D.} \bibnamefont{de~R{\'e}otier}},
  \bibinfo{author}{\bibfnamefont{A.}~\bibnamefont{Yaouanc}},
  \bibinfo{author}{\bibfnamefont{C.}~\bibnamefont{Baines}},
  \bibinfo{author}{\bibfnamefont{G.}~\bibnamefont{Lapertot}},
  \bibinfo{author}{\bibfnamefont{K.}~\bibnamefont{Mony}},
  \bibinfo{author}{\bibfnamefont{J.-P.} \bibnamefont{Brison}},
  \bibnamefont{and} \bibinfo{author}{\bibfnamefont{H.}~\bibnamefont{Keller}},
  \bibinfo{journal}{Physical Review Letters} \textbf{\bibinfo{volume}{110}},
  \bibinfo{pages}{017005} (\bibinfo{year}{2013}).

\bibitem[{\citenamefont{Cho et~al.}(2016)\citenamefont{Cho, Ko{\'n}czykowski,
  Teknowijoyo, Tanatar, Liu, Lograsso, Straszheim, Mishra, Maiti, Hirschfeld
  et~al.}}]{Cho2016}
\bibinfo{author}{\bibfnamefont{K.}~\bibnamefont{Cho}},
  \bibinfo{author}{\bibfnamefont{M.}~\bibnamefont{Ko{\'n}czykowski}},
  \bibinfo{author}{\bibfnamefont{S.}~\bibnamefont{Teknowijoyo}},
  \bibinfo{author}{\bibfnamefont{M.~A.} \bibnamefont{Tanatar}},
  \bibinfo{author}{\bibfnamefont{Y.}~\bibnamefont{Liu}},
  \bibinfo{author}{\bibfnamefont{T.~A.} \bibnamefont{Lograsso}},
  \bibinfo{author}{\bibfnamefont{W.~E.} \bibnamefont{Straszheim}},
  \bibinfo{author}{\bibfnamefont{V.}~\bibnamefont{Mishra}},
  \bibinfo{author}{\bibfnamefont{S.}~\bibnamefont{Maiti}},
  \bibinfo{author}{\bibfnamefont{P.~J.} \bibnamefont{Hirschfeld}},
  \bibnamefont{et~al.}, \bibinfo{journal}{Science Advances}
  \textbf{\bibinfo{volume}{2}}, \bibinfo{pages}{e1600807}
  (\bibinfo{year}{2016}).

\bibitem[{\citenamefont{Kawano-Furukawa
  et~al.}(2013)\citenamefont{Kawano-Furukawa, DeBeer-Schmitt, Kikuchi, Cameron,
  Holmes, Heslop, Forgan, White, Kihou, Lee et~al.}}]{Kawano2013}
\bibinfo{author}{\bibfnamefont{H.}~\bibnamefont{Kawano-Furukawa}},
  \bibinfo{author}{\bibfnamefont{L.}~\bibnamefont{DeBeer-Schmitt}},
  \bibinfo{author}{\bibfnamefont{H.}~\bibnamefont{Kikuchi}},
  \bibinfo{author}{\bibfnamefont{A.~S.} \bibnamefont{Cameron}},
  \bibinfo{author}{\bibfnamefont{A.~T.} \bibnamefont{Holmes}},
  \bibinfo{author}{\bibfnamefont{R.~W.} \bibnamefont{Heslop}},
  \bibinfo{author}{\bibfnamefont{E.~M.} \bibnamefont{Forgan}},
  \bibinfo{author}{\bibfnamefont{J.~S.} \bibnamefont{White}},
  \bibinfo{author}{\bibfnamefont{K.}~\bibnamefont{Kihou}},
  \bibinfo{author}{\bibfnamefont{C.~H.} \bibnamefont{Lee}},
  \bibnamefont{et~al.}, \bibinfo{journal}{Physical Review B}
  \textbf{\bibinfo{volume}{88}}, \bibinfo{pages}{134524}
  (\bibinfo{year}{2013}).

\bibitem[{\citenamefont{Putzke et~al.}({2014})\citenamefont{Putzke, Walmsley,
  Fletcher, Malone, Vignolles, Proust, Badoux, See, Beere, Ritchie
  et~al.}}]{Putzke2014}
\bibinfo{author}{\bibfnamefont{C.}~\bibnamefont{Putzke}},
  \bibinfo{author}{\bibfnamefont{P.}~\bibnamefont{Walmsley}},
  \bibinfo{author}{\bibfnamefont{J.~D.} \bibnamefont{Fletcher}},
  \bibinfo{author}{\bibfnamefont{L.}~\bibnamefont{Malone}},
  \bibinfo{author}{\bibfnamefont{D.}~\bibnamefont{Vignolles}},
  \bibinfo{author}{\bibfnamefont{C.}~\bibnamefont{Proust}},
  \bibinfo{author}{\bibfnamefont{S.}~\bibnamefont{Badoux}},
  \bibinfo{author}{\bibfnamefont{P.}~\bibnamefont{See}},
  \bibinfo{author}{\bibfnamefont{H.~E.} \bibnamefont{Beere}},
  \bibinfo{author}{\bibfnamefont{D.~A.} \bibnamefont{Ritchie}},
  \bibnamefont{et~al.}, \bibinfo{journal}{{Nat. Commun.}}
  \textbf{\bibinfo{volume}{{5}}}, \bibinfo{pages}{{5679}}
  (\bibinfo{year}{{2014}}).

\bibitem[{\citenamefont{Brandt}(1999)}]{Brandt1999}
\bibinfo{author}{\bibfnamefont{E.~H.} \bibnamefont{Brandt}},
  \bibinfo{journal}{Phys. Rev. B} \textbf{\bibinfo{volume}{60}},
  \bibinfo{pages}{11939} (\bibinfo{year}{1999}).

\bibitem[{\citenamefont{Majumdar et~al.}(2003)\citenamefont{Majumdar, Lees,
  Balakrishnan, and Paul}}]{Majumdar2003}
\bibinfo{author}{\bibfnamefont{S.}~\bibnamefont{Majumdar}},
  \bibinfo{author}{\bibfnamefont{M.~R.} \bibnamefont{Lees}},
  \bibinfo{author}{\bibfnamefont{G.}~\bibnamefont{Balakrishnan}},
  \bibnamefont{and} \bibinfo{author}{\bibfnamefont{D.~M.} \bibnamefont{Paul}},
  \bibinfo{journal}{Physical Review B} \textbf{\bibinfo{volume}{68}}
  (\bibinfo{year}{2003}).

\bibitem[{\citenamefont{Cooper}(1996)}]{Cooper1996}
\bibinfo{author}{\bibfnamefont{J.~R.} \bibnamefont{Cooper}},
  \bibinfo{journal}{Phys. Rev. B} \textbf{\bibinfo{volume}{54}},
  \bibinfo{pages}{R3753} (\bibinfo{year}{1996}).

\bibitem[{\citenamefont{Serafin et~al.}(2010)\citenamefont{Serafin, Coldea,
  Ganin, Rosseinsky, Prassides, Vignolles, and Carrington}}]{Serafin2010}
\bibinfo{author}{\bibfnamefont{A.}~\bibnamefont{Serafin}},
  \bibinfo{author}{\bibfnamefont{A.~I.} \bibnamefont{Coldea}},
  \bibinfo{author}{\bibfnamefont{A.~Y.} \bibnamefont{Ganin}},
  \bibinfo{author}{\bibfnamefont{M.~J.} \bibnamefont{Rosseinsky}},
  \bibinfo{author}{\bibfnamefont{K.}~\bibnamefont{Prassides}},
  \bibinfo{author}{\bibfnamefont{D.}~\bibnamefont{Vignolles}},
  \bibnamefont{and}
  \bibinfo{author}{\bibfnamefont{A.}~\bibnamefont{Carrington}},
  \bibinfo{journal}{Phys. Rev. B} \textbf{\bibinfo{volume}{82}},
  \bibinfo{pages}{104514} (\bibinfo{year}{2010}).

\bibitem[{\citenamefont{Hu}(1994)}]{Hu1994}
\bibinfo{author}{\bibfnamefont{C.-R.} \bibnamefont{Hu}},
  \bibinfo{journal}{Physical review letters} \textbf{\bibinfo{volume}{72}},
  \bibinfo{pages}{1526} (\bibinfo{year}{1994}).

\bibitem[{\citenamefont{Fogelstr{\"o}m
  et~al.}(1997)\citenamefont{Fogelstr{\"o}m, Rainer, and
  Sauls}}]{Fogelstrom1997}
\bibinfo{author}{\bibfnamefont{M.}~\bibnamefont{Fogelstr{\"o}m}},
  \bibinfo{author}{\bibfnamefont{D.}~\bibnamefont{Rainer}}, \bibnamefont{and}
  \bibinfo{author}{\bibfnamefont{J.~A.} \bibnamefont{Sauls}},
  \bibinfo{journal}{Physical review letters} \textbf{\bibinfo{volume}{79}},
  \bibinfo{pages}{281} (\bibinfo{year}{1997}).

\bibitem[{\citenamefont{Barash et~al.}(2000)\citenamefont{Barash, Kalenkov, and
  Kurkij{\"a}rvi}}]{Barash2000}
\bibinfo{author}{\bibfnamefont{Y.~S.} \bibnamefont{Barash}},
  \bibinfo{author}{\bibfnamefont{M.}~\bibnamefont{Kalenkov}}, \bibnamefont{and}
  \bibinfo{author}{\bibfnamefont{J.}~\bibnamefont{Kurkij{\"a}rvi}},
  \bibinfo{journal}{Physical Review B} \textbf{\bibinfo{volume}{62}},
  \bibinfo{pages}{6665} (\bibinfo{year}{2000}).

\bibitem[{\citenamefont{Walter et~al.}(1998)\citenamefont{Walter, Prusseit,
  Semerad, Kinder, Assmann, Huber, Burkhardt, Rainer, and Sauls}}]{Walter1998}
\bibinfo{author}{\bibfnamefont{H.}~\bibnamefont{Walter}},
  \bibinfo{author}{\bibfnamefont{W.}~\bibnamefont{Prusseit}},
  \bibinfo{author}{\bibfnamefont{R.}~\bibnamefont{Semerad}},
  \bibinfo{author}{\bibfnamefont{H.}~\bibnamefont{Kinder}},
  \bibinfo{author}{\bibfnamefont{W.}~\bibnamefont{Assmann}},
  \bibinfo{author}{\bibfnamefont{H.}~\bibnamefont{Huber}},
  \bibinfo{author}{\bibfnamefont{H.}~\bibnamefont{Burkhardt}},
  \bibinfo{author}{\bibfnamefont{D.}~\bibnamefont{Rainer}}, \bibnamefont{and}
  \bibinfo{author}{\bibfnamefont{J.~A.} \bibnamefont{Sauls}},
  \bibinfo{journal}{Physical Review Letters} \textbf{\bibinfo{volume}{80}},
  \bibinfo{pages}{3598} (\bibinfo{year}{1998}).

\bibitem[{\citenamefont{Carrington et~al.}(2001)\citenamefont{Carrington,
  Manzano, Prozorov, Giannetta, Kameda, and Tamegai}}]{Carrington2001}
\bibinfo{author}{\bibfnamefont{A.}~\bibnamefont{Carrington}},
  \bibinfo{author}{\bibfnamefont{F.}~\bibnamefont{Manzano}},
  \bibinfo{author}{\bibfnamefont{R.}~\bibnamefont{Prozorov}},
  \bibinfo{author}{\bibfnamefont{R.~W.} \bibnamefont{Giannetta}},
  \bibinfo{author}{\bibfnamefont{N.}~\bibnamefont{Kameda}}, \bibnamefont{and}
  \bibinfo{author}{\bibfnamefont{T.}~\bibnamefont{Tamegai}},
  \bibinfo{journal}{Phys. Rev. Lett.} \textbf{\bibinfo{volume}{86}},
  \bibinfo{pages}{1074} (\bibinfo{year}{2001}).

\bibitem[{\citenamefont{Ara{\'u}jo and Sacramento}(2009)}]{Araujo2009}
\bibinfo{author}{\bibfnamefont{M.}~\bibnamefont{Ara{\'u}jo}} \bibnamefont{and}
  \bibinfo{author}{\bibfnamefont{P.}~\bibnamefont{Sacramento}},
  \bibinfo{journal}{Physical Review B} \textbf{\bibinfo{volume}{79}},
  \bibinfo{pages}{174529} (\bibinfo{year}{2009}).

\bibitem[{\citenamefont{Huang and Lin}(2010)}]{Huang2010}
\bibinfo{author}{\bibfnamefont{W.-M.} \bibnamefont{Huang}} \bibnamefont{and}
  \bibinfo{author}{\bibfnamefont{H.-H.} \bibnamefont{Lin}},
  \bibinfo{journal}{Physical Review B} \textbf{\bibinfo{volume}{81}},
  \bibinfo{pages}{052504} (\bibinfo{year}{2010}).

\bibitem[{\citenamefont{Fletcher et~al.}(2009)\citenamefont{Fletcher, Serafin,
  Malone, Analytis, Chu, Erickson, Fisher, and Carrington}}]{Fletcher2009}
\bibinfo{author}{\bibfnamefont{J.}~\bibnamefont{Fletcher}},
  \bibinfo{author}{\bibfnamefont{A.}~\bibnamefont{Serafin}},
  \bibinfo{author}{\bibfnamefont{L.}~\bibnamefont{Malone}},
  \bibinfo{author}{\bibfnamefont{J.}~\bibnamefont{Analytis}},
  \bibinfo{author}{\bibfnamefont{J.-H.} \bibnamefont{Chu}},
  \bibinfo{author}{\bibfnamefont{A.}~\bibnamefont{Erickson}},
  \bibinfo{author}{\bibfnamefont{I.}~\bibnamefont{Fisher}}, \bibnamefont{and}
  \bibinfo{author}{\bibfnamefont{A.}~\bibnamefont{Carrington}},
  \bibinfo{journal}{Phys. Rev. Lett.} \textbf{\bibinfo{volume}{102}},
  \bibinfo{pages}{147001} (\bibinfo{year}{2009}).

\bibitem[{\citenamefont{Stojkovi{\'c} and Valls}(1995)}]{Stojkovic1995}
\bibinfo{author}{\bibfnamefont{B.~P.} \bibnamefont{Stojkovi{\'c}}}
  \bibnamefont{and} \bibinfo{author}{\bibfnamefont{O.~T.} \bibnamefont{Valls}},
  \bibinfo{journal}{Phys. Rev. B.} \textbf{\bibinfo{volume}{51}},
  \bibinfo{pages}{6049} (\bibinfo{year}{1995}).

\bibitem[{\citenamefont{Xu et~al.}(1995)\citenamefont{Xu, Yip, and
  Sauls}}]{Xu1995}
\bibinfo{author}{\bibfnamefont{D.}~\bibnamefont{Xu}},
  \bibinfo{author}{\bibfnamefont{S.~K.} \bibnamefont{Yip}}, \bibnamefont{and}
  \bibinfo{author}{\bibfnamefont{J.~A.} \bibnamefont{Sauls}},
  \bibinfo{journal}{Phys. Rev. B} \textbf{\bibinfo{volume}{51}},
  \bibinfo{pages}{16233} (\bibinfo{year}{1995}).

\bibitem[{\citenamefont{Deepwell et~al.}(2013)\citenamefont{Deepwell, Peets,
  Truncik, Murphy, Kennett, Huttema, Liang, Bonn, Hardy, and
  Broun}}]{Deepwell2013}
\bibinfo{author}{\bibfnamefont{D.}~\bibnamefont{Deepwell}},
  \bibinfo{author}{\bibfnamefont{D.~C.} \bibnamefont{Peets}},
  \bibinfo{author}{\bibfnamefont{C.~J.~S.} \bibnamefont{Truncik}},
  \bibinfo{author}{\bibfnamefont{N.~C.} \bibnamefont{Murphy}},
  \bibinfo{author}{\bibfnamefont{M.~P.} \bibnamefont{Kennett}},
  \bibinfo{author}{\bibfnamefont{W.~A.} \bibnamefont{Huttema}},
  \bibinfo{author}{\bibfnamefont{R.}~\bibnamefont{Liang}},
  \bibinfo{author}{\bibfnamefont{D.~A.} \bibnamefont{Bonn}},
  \bibinfo{author}{\bibfnamefont{W.~N.} \bibnamefont{Hardy}}, \bibnamefont{and}
  \bibinfo{author}{\bibfnamefont{D.~M.} \bibnamefont{Broun}},
  \bibinfo{journal}{Phys. Rev. B} \textbf{\bibinfo{volume}{88}},
  \bibinfo{pages}{214509} (\bibinfo{year}{2013}).

\bibitem[{\citenamefont{Prozorov and Giannetta}(2006)}]{Prozorov2006}
\bibinfo{author}{\bibfnamefont{R.}~\bibnamefont{Prozorov}} \bibnamefont{and}
  \bibinfo{author}{\bibfnamefont{R.~W.} \bibnamefont{Giannetta}},
  \bibinfo{journal}{Supercond. Sci. Tech.} \textbf{\bibinfo{volume}{19}},
  \bibinfo{pages}{R41} (\bibinfo{year}{2006}), ISSN \bibinfo{issn}{0953-2048,
  1361-6668}.

\bibitem[{\citenamefont{Uemura}(2009)}]{Uemura2009}
\bibinfo{author}{\bibfnamefont{Y.}~\bibnamefont{Uemura}},
  \bibinfo{journal}{Physica B} \textbf{\bibinfo{volume}{404}},
  \bibinfo{pages}{3195 } (\bibinfo{year}{2009}).

\bibitem[{\citenamefont{Ormeno et~al.}(2002)\citenamefont{Ormeno, Sibley,
  Gough, Sebastian, and Fisher}}]{Ormeno2002}
\bibinfo{author}{\bibfnamefont{R.~J.} \bibnamefont{Ormeno}},
  \bibinfo{author}{\bibfnamefont{A.}~\bibnamefont{Sibley}},
  \bibinfo{author}{\bibfnamefont{C.~E.} \bibnamefont{Gough}},
  \bibinfo{author}{\bibfnamefont{S.}~\bibnamefont{Sebastian}},
  \bibnamefont{and} \bibinfo{author}{\bibfnamefont{I.~R.}
  \bibnamefont{Fisher}}, \bibinfo{journal}{Phys. Rev. Lett.}
  \textbf{\bibinfo{volume}{88}} (\bibinfo{year}{2002}).

\bibitem[{\citenamefont{Coldea et~al.}(2008)\citenamefont{Coldea, Fletcher,
  Carrington, Analytis, Bangura, Chu, Erickson, Fisher, Hussey, and
  McDonald}}]{Coldea2008}
\bibinfo{author}{\bibfnamefont{A.~I.} \bibnamefont{Coldea}},
  \bibinfo{author}{\bibfnamefont{J.~D.} \bibnamefont{Fletcher}},
  \bibinfo{author}{\bibfnamefont{A.}~\bibnamefont{Carrington}},
  \bibinfo{author}{\bibfnamefont{J.~G.} \bibnamefont{Analytis}},
  \bibinfo{author}{\bibfnamefont{A.~F.} \bibnamefont{Bangura}},
  \bibinfo{author}{\bibfnamefont{J.-H.} \bibnamefont{Chu}},
  \bibinfo{author}{\bibfnamefont{A.~S.} \bibnamefont{Erickson}},
  \bibinfo{author}{\bibfnamefont{I.~R.} \bibnamefont{Fisher}},
  \bibinfo{author}{\bibfnamefont{N.~E.} \bibnamefont{Hussey}},
  \bibnamefont{and} \bibinfo{author}{\bibfnamefont{R.~D.}
  \bibnamefont{McDonald}}, \bibinfo{journal}{Phys. Rev. Lett.}
  \textbf{\bibinfo{volume}{101}}, \bibinfo{pages}{216402}
  (\bibinfo{year}{2008}).

\bibitem[{\citenamefont{Carrington et~al.}(2009)\citenamefont{Carrington,
  Coldea, Fletcher, Hussey, Andrew, Bangura, Analytis, Chu, Erickson, Fisher
  et~al.}}]{carrington2009}
\bibinfo{author}{\bibfnamefont{A.}~\bibnamefont{Carrington}},
  \bibinfo{author}{\bibfnamefont{A.}~\bibnamefont{Coldea}},
  \bibinfo{author}{\bibfnamefont{J.}~\bibnamefont{Fletcher}},
  \bibinfo{author}{\bibfnamefont{N.}~\bibnamefont{Hussey}},
  \bibinfo{author}{\bibfnamefont{C.}~\bibnamefont{Andrew}},
  \bibinfo{author}{\bibfnamefont{A.}~\bibnamefont{Bangura}},
  \bibinfo{author}{\bibfnamefont{J.}~\bibnamefont{Analytis}},
  \bibinfo{author}{\bibfnamefont{J.-H.} \bibnamefont{Chu}},
  \bibinfo{author}{\bibfnamefont{A.}~\bibnamefont{Erickson}},
  \bibinfo{author}{\bibfnamefont{I.}~\bibnamefont{Fisher}},
  \bibnamefont{et~al.}, \bibinfo{journal}{Physica C}
  \textbf{\bibinfo{volume}{469}}, \bibinfo{pages}{459 } (\bibinfo{year}{2009}).

\bibitem[{\citenamefont{Settai et~al.}(2001)\citenamefont{Settai, Shishido,
  Ikeda, Murakawa, Nakashima, Aoki, Haga, Harima, and Onuki}}]{Settai2001}
\bibinfo{author}{\bibfnamefont{R.}~\bibnamefont{Settai}},
  \bibinfo{author}{\bibfnamefont{H.}~\bibnamefont{Shishido}},
  \bibinfo{author}{\bibfnamefont{S.}~\bibnamefont{Ikeda}},
  \bibinfo{author}{\bibfnamefont{Y.}~\bibnamefont{Murakawa}},
  \bibinfo{author}{\bibfnamefont{M.}~\bibnamefont{Nakashima}},
  \bibinfo{author}{\bibfnamefont{D.}~\bibnamefont{Aoki}},
  \bibinfo{author}{\bibfnamefont{Y.}~\bibnamefont{Haga}},
  \bibinfo{author}{\bibfnamefont{H.}~\bibnamefont{Harima}}, \bibnamefont{and}
  \bibinfo{author}{\bibfnamefont{Y.}~\bibnamefont{Onuki}}, \bibinfo{journal}{J.
  Phys. Cond. Matt.} \textbf{\bibinfo{volume}{13}}, \bibinfo{pages}{L627}
  (\bibinfo{year}{2001}).

\bibitem[{\citenamefont{Miyake and Varma}(2018)}]{Miyake2018}
\bibinfo{author}{\bibfnamefont{K.}~\bibnamefont{Miyake}} \bibnamefont{and}
  \bibinfo{author}{\bibfnamefont{C.~M.} \bibnamefont{Varma}},
  \bibinfo{journal}{Phys. Rev. B} \textbf{\bibinfo{volume}{98}},
  \bibinfo{pages}{174501} (\bibinfo{year}{2018}).

\end{thebibliography}


\begin{thebibliography}{36}

\bibitem{Hirschfeld2011} Hirschfeld, P., Korshunov, M., Mazin, I. Gap symmetry and structure of Fe-based superconductors. \textit{Rep. Prog. Phys.} \textbf{74}, 124508 (2011).

\bibitem{Prozorov2006} Prozorov, R., Giannetta, R.W. Magnetic penetration depth in unconventional superconductors. \textit{Supercond. Sci. Tech.} \textbf{19}, R41 (2006).

\bibitem{Carrington2011} Carrington, A. Studies of the gap structure of iron-based superconductors using magnetic penetration depth. \textit{Comptes Rendus Physique} \textbf{12}, 502 (2011).

\bibitem{Yip1992}  Yip, S.K., Sauls, J.A. Nonlinear Meissner effect in CuO Superconductors. \textit{Phys. Rev. Lett.} \textbf{69}, 2264 (1992).

\bibitem{Carrington1999} Carrington, A., Giannetta, R.W., Kim, J.T., Giapintzakis, J. Absence of nonlinear Meissner effect in YBa$_2$Cu$_3$O$_{6.95}$, \textit{Phys. Rev. B} \textbf{59}, R14173 (1999).

\bibitem{Bidinosti1999} Bidinosti, C.P., Hardy, W.N., Bonn, D.A., Liang, R. Magnetic field dependence of $\lambda$ in YBa$_2$Cu$_3$O$_{6.95}$: results as a function of temperature and field orientation. \textit{Phys. Rev. Lett.} \textbf{83}, 3277 (1999).

\bibitem{Halterman2001} Halterman, K., Valls, O.T., \v{Z}uti\'{c} I. Reanalysis of the magnetic field dependence of the penetration depth: Observation of the nonlinear Meissner effect \textit{Phys. Rev. B} \textbf{63}, 180405R (2001).


\bibitem{Bhattacharya1999} Bhattacharya, A. \textit{et.\ al}. Angular dependence of the nonlinear transverse magnetic moment of YBa$_2$Cu$_3$O$_{6.95}$ in the Meissner state, \textit{Phys. Rev. Lett.} \textbf{82}, 3132 (1999).

\bibitem{Dahm97} Dahm, T. and Scalapino, D.J. Temperature dependence of intermodulation distortion in YBCO \textit{J. Appl. Phys.} \textbf{81}, 2002 (1997).

\bibitem{Oates2004} Oates, D.E., Park, S.-H., Koren, G. Observation of the nonlinear Meissner Effect in YBCO thin films: evidence for a $d$-wave order parameter in the bulk of the cuprate superconductors, \textit{Phys. Rev. Lett.} \textbf{93}, 197001 (2004).

\bibitem{Movshovich2001} Movshovich, R. \textit{et.\ al}. Unconventional superconductivity in CeIrIn$_5$ and CeCoIn$_5$ : specific Heat and thermal conductivity studies, \textit{Phys. Rev. Lett.} \textbf{86}, 5152 (2001).

\bibitem{Seyfarth2008} Seyfarth, G. \textit{et.\ al}. Multigap superconductivity in the heavy-fermion system CeCoIn$_5$, \textit{Phys. Rev. Lett.} \textbf{101}, 046401 (2008).

\bibitem{Ormeno2002} Ormeno, R.J., Sibley, A., Gough, C.E., Sebastian, S., Fisher, I.R. Microwave conductivity and penetration depth in the heavy fermion superconductor CeCoIn$_5$, \textit{Phys. Rev. Lett.} \textbf{88}, (2002).

\bibitem{Chia2003} Chia, E.E. \textit{et.\ al}. Nonlocality and strong coupling in the heavy fermion superconductor CeCoIn$_5$: a penetration depth study, \textit{Phys. Rev. B} \textbf{67}, 014527 (2003).

\bibitem{Ozcan2003} \"{O}zcan, S. \textit{et.\ al}. London penetration depth measurements of the heavy-fermion superconductor CeCoIn$_5$ near a magnetic quantum critical point, \textit{Europhys. Lett.} \textbf{62}, 412 (2003).

\bibitem{Fletcher2009} Fletcher, J. \textit{et.\ al}. Evidence for a nodal-line superconducting state in LaFePO, \textit{Phys. Rev. Lett.} \textbf{102}, 147001 (2009).

\bibitem{Hicks2009} Hicks, C.W. \textit{et.\ al}. Evidence for a nodal energy gap in the iron-pnictide superconductor LaFePO from penetration depth measurements by scanning squid susceptometry, \textit{Phys. Rev. Lett.} \textbf{103}, 127003 (2009).

\bibitem{Yamashita2009} Yamashita, M. \textit{et.\ al}. Thermal conductivity measurements of the energy-gap anisotropy of superconducting LaFePO at low temperatures, \textit{Phys. Rev. B} \textbf{80}, 220509 (2009).

\bibitem{Sutherland2012} Sutherland, M. \textit{et.\ al}. Low-energy quasiparticles probed by heat transport in the iron-based superconductor LaFePO, \textit{Phys. Rev. B} \textit{85}, 014517 (2012).

\bibitem{Kuroki2009} Kuroki, K., Usui, H., Onari, S., Arita, R., Aoki, H. Pnictogen height as a possible switch between high-$T_c$ nodeless and low-$T_c$ nodal pairings in the iron-based superconductors, \textit{Phys. Rev. B} \textbf{79}, 224511 (2009).

\bibitem{Hashimoto2010} Hashimoto, K. \textit{et.\ al}. Evidence for superconducting gap nodes in the zone-centered hole bands of KFe$_2$As$_2$ from magnetic penetration-depth measurements, \textit{Phys. Rev. B} \textbf{82}, 014526 (2010).

\bibitem{Hardy2013} Hardy, F. \textit{et.\ al}. Multiband superconductivity in KFe$_2$As$_2$: Evidence for one isotropic and several Lilliputian energy gaps, \textit{J. Phys. Soc. Jap.} \textbf{83}, 014711 (2013).

\bibitem{Reid2012} Reid, J.-P. \textit{et.\ al}. Universal heat conduction in the iron arsenide superconductor KFe$_2$As$_2$: evidence of a $d$-wave state, \textit{Phys. Rev. Lett.} \textbf{109}, 087001 (2012).

\bibitem{Watanabe2014} Watanabe, D. \textit{et.\ al}. Doping evolution of the quasiparticle excitations in heavily hole-doped Ba$_{1-x}$K$_x$Fe$_2$As$_2$: a possible superconducting gap with sign-reversal between hole pockets, \textit{Phys. Rev. B} \textbf{89}, 115112 (2014).

\bibitem{Okazaki2012} Okazaki, K. \textit{et.\ al}. Octet-line node structure of superconducting order parameter in KFe$_2$As$_2$, \textit{Science} \textbf{337}, 1314 (2012)

\bibitem{Cho2016} Cho, K. \textit{et.\ al}. Energy gap evolution across the superconductivity dome in single crystals of (Ba$_{1-x}$K$_x$)Fe$_2$As$_2$, \textit{Sci. Adv.} \textbf{2}, e1600807 (2016).

\bibitem{Hashimoto2013} Hashimoto, K. \textit{et.\ al}. Anomalous superfluid density in quantum critical superconductors, \textit{Proc. Natl. Acad. Sci.} \textbf{110}, 3293 (2013).

\bibitem{hirschfeld1993}  Hirschfeld, P.J.,  Goldenfeld, N. Effect of strong scattering on the low-temperature penetration depth of a $d$-wave superconductor. \textit{Phys. Rev. B.} \textbf{48}, 4219 (1993).

\bibitem{Cooper1996}  Cooper, J.R. Power-law dependence of the |$ab$-plane penetration depth in Nd$_{1.85}$Ce$_{0.15}$CuO$_{4-y}$, \textit{Phys. Rev. B} \textbf{54}, R3753 (1996).

\bibitem{Xu1995}  Xu, D., Yip, S.K., Sauls, J.A. Nonlinear Meissner effect in unconventional superconductors. \textit{Phys. Rev. B} \textbf{51}, 16233 (1995).

\bibitem{Stojkovic1995} Stojkovi\'{c}, B.P., Valls, O.T. Nonlinear supercurrent response in anisotropic superconductors. \textit{Phys.Rev. B.} \textbf{51}, 6049 (1995).

\bibitem{Carrington2001} Carrington, A. \textit{et.\ al}. Evidence for surface Andreev bound states in cuprate superconductors from penetration depth measurements, \textit{Phys. Rev. Lett.} \textbf{86}, 1074 (2001).

\bibitem{Barash2000} Barash, Y.S., Kalenkov, M., Kurkij\"{a}rvi, J. Low-temperature magnetic penetration depth in $d$-wave superconductors: Zero-energy bound state and impurity effects, \textit{Phys. Rev. B} \textbf{62}, 6665 (2000).

\bibitem{Zare2010} Zare, A., Dahm, T., Schopohl, N. Strong surface contribution to the nonlinear Meissner effect in $d$-wave superconductors, \textit{Phys. Rev. Lett.} \textbf{104}, 237001 (2010).

\bibitem{Araujol2009} Ara\'{u}jol, M.A.N. Sacramento, P.D. Quantum waveguide theory of Andreev spectroscopy in multiband superconductors:The case of iron pnictides \textit{Phys. Rev. B} \textbf{79}, 174529 (2009).

\bibitem{Wilcox2019} Details of this study will be published separately.

\bibitem{Uemura2009} Uemura, Y.J. Energy-scale phenomenology and pairing via resonant spin-charge motion in FeAs CuO, heavy-fermion and other exotic superconductors, \textit{Physica B} \textbf{404}, 3195 (2009).

\bibitem{Furukawa2011} Kawano-Furukawa, H. \textit{et.\ al}. Gap in KFe$_2$As$_2$ studied by small-angle neutron scattering observations of the magnetic vortex lattice \textit{Phys. Rev. B} \textbf{84}, 024507 (2011).

%\bibitem{Pippard1950} Pippard, A. Field variation of the superconducting penetration depth. \textit{Proc. Roy. Soc. London.}  \textbf{A203}, 210 (1950).

%\bibitem{Benz2001} Benz, G., W\"{u}nsch, S., Scherer, T., Neuhaus, M., Jutzi, W. Measured temperature dependence of the intermodulation product of coplanar waveguides with $s$- and $d$-wave superconductors, \textit{Physica C: Superconductivity} \textbf{356}, 122 (2001).

%\bibitem{Zhuravel2013} Zhuravel, A.P. \textit{et.\ al}. Imaging the anisotropic nonlinear Meissner effect in nodal YBa$_2$Cu$_3$O$_7$ thin-film superconductors, \textit{Phys. Rev. Lett.} \textbf{110}, 087002 (2013).

%\bibitem{Movshovich2001} Movshovich, R., Jaime, M., Thompson, J.D., Petrovic, C., Fisk, Z., Pagliuso, P.G., Sarrao, J.L. Unconventional Superconductivity in CeIrIn$_5$ and CeCoIn$_5$ : Specific Heat and Thermal Conductivity Studies, \textit{Phys. Rev. Lett.} \textbf{86}, 5152 (2001).

%bibitem{Seyfarth2008} Seyfarth, G., Brison, J.P., Knebel, G., Aoki, D., Lapertot, G., Flouquet, J. Multigap superconductivity in the heavy-fermion system CeCoIn$_5$, \textit{Phys. Rev. Lett.} \textbf{101}, 046401 (2008).

%\bibitem{Chia2003} Chia, E.E., Van Harlingen, D., Salamon, M., Yanoff, B.D., Bonalde, I., Sarrao, J. Nonlocality and Strong Coupling in the Heavy Fermion Superconductor CeCoIn$_5$: A Penetration Depth Study, \textit{Phys. Rev. B} \textbf{67}, 014527 (2003).

%\bibitem{Ozcan2003} \"{O}zcan, S., Broun, D.M., Morgan, B., Haselwimmer, R.K.W., Sarrao, J.L., Kamal, S., Bidinosti, C.P., Turner, P.J., Raudsepp, M., Waldram, J.R. London Penetration Depth Measurements of the Heavy-Fermion Superconductor CeCoIn$_5$ near a Magnetic Quantum Critical Point, \textit{Europhys. Lett.} \textbf{62}, 412 (2003).

%\bibitem{Fletcher2009} Fletcher, J., Serafin, A., Malone, L., Analytis, J., Chu, J.-H., Erickson, A., Fisher, I., Carrington, A. Evidence for a nodal-line superconducting state in LaFePO, \textit{Phys. Rev. Lett.} \textbf{102}, 147001 (2009).

%\bibitem{Hicks2009} Hicks, C.W., Lippman, T.M., Huber, M.E., Analytis, J.G., Chu, J.-H., Erickson, A.S., Fisher, I.R., Moler, K.A. Evidence for a nodal energy gap in the iron-pnictide superconductor LaFePO from penetration depth measurements by scanning squid susceptometry, \textit{Phys. Rev. Lett.} \textbf{103}, 127003 (2009).

%\bibitem{Yamashita2009} Yamashita, M., Nakata, N., Senshu, Y., Tonegawa, S., Ikada, K., Hashimoto, K., Sugawara, H., Shibauchi, T., Matsuda, Y. Thermal conductivity measurements of the energy-gap anisotropy of superconducting LaFePO at low temperatures, \textit{Phys. Rev. B} \textbf{80}, 220509 (2009).

%\bibitem{Sutherland2012} Sutherland, M., Dunn, J., Toews, W.H., O'Farrell, E., Analytis, J., Fisher, I., Hill, R.W. Low-energy quasiparticles probed by heat transport in the iron-based superconductor LaFePO, \textit{Phys. Rev. B} \textit{85}, 014517 (2012).

%\bibitem{Hashimoto2010} Hashimoto, K., Serafin, A., Tonegawa, S., Katsumata, R., Okazaki, R., Saito, T., Fukazawa, H., Kohori, Y., Kihou, K., Lee, C.H, Iyo, A., Eisaki, H., Ikeda, H., Matsuda, Y., Carrington, A., Shibauchi, T. Evidence for superconducting gap nodes in the zone-centered hole bands of KFe$_2$As$_2$ from magnetic penetration-depth measurements, \textit{Phys. Rev. B} \textbf{82}, 014526 (2010).

%\bibitem{Hardy2013} Hardy, F., Eder, R., Jackson, M., Aoki, D., Paulsen, C., Wolf, T., Burger, P., B\"{o}hmer, A., Schweiss, P., Adelmann,  P., Fisher, R.A., Meingast, C. Multiband superconductivity in KFe$_2$As$_2$: Evidence for one isotropic and several lilliputian energy gaps, \textit{J. Phys. Soc. Jap.} \textbf{83}, 014711 (2013).

%\bibitem{Reid2012} Reid, J.-P., Tanatar, M.A., Juneau-Fecteau, A., Gordon, R., de Cotret, S.R., Doiron-Leyraud, N., Saito, T., Fukazawa, H., Kohori, Y., Kihou, K., Lee, C.H., Iyo, A., Eisaki, H., Prozorov, R., Taillefer, L. Universal heat conduction in the iron arsenide superconductor KFe$_2$As$_2$: evidence of a $d$-wave state, \textit{Phys. Rev. Lett.} \textbf{109}, 087001 (2012).

%\bibitem{Watanabe2014} Watanabe, D., Yamashita, T., Kawamoto, Y., Kurata, S., Mizukami, Y., Ohta, T., Kasahara, S., Yamashita, M., Saito, T., Fukazawa, H., Kohori, Y., Ishida, S., Kihou, K., Lee, C.H., Iyo, A., Eisaki, H., Vorontsov, A.B., Shibauchi, T., Matsuda, Y. Doping evolution of the quasiparticle excitations in heavily hole-doped Ba$_{1-x}$K$_x$Fe$_2$As$_2$: A possible superconducting gap with sign-reversal between hole pockets, \textit{Phys. Rev. B} \textbf{89}, 115112 (2014).

%\bibitem{Okazaki2012} Okazaki, K., Ota, Y., Kotani, Y., Malaeb, W., Ishida, Y., Shimojima, T., Kiss, T., Watanabe, S., Chen, C.-T., Kihou, K., Lee, C.H., Iyo, A., Eisaki, H., Saito, T., Fukazawa, H., Korohi, Y., Hashimoto, K., Shibauchi, T., Matsuda,, Y., Ikeda, H., Miyahara, H., Arita, R., Chainani, A., Shin, S. Octet-line node structure of superconducting order parameter in KFe$_2$As$_2$, \textit{Science} \textbf{337}, 1314 (2012)

%\bibitem{Cho2016} Cho, K., K\'{o}nczykowski, M., Teknowijoyo, S., Tanatar, M.A., Liu, Y., Lograsso, T.A., Straszheim, W.E., Mishra, V., Maiti, S., Hirschfeld, P.J., Prozorov, R. Energy gap evolution across the superconductivity dome in single crystals of (Ba$_{1-x}$K$_x$)Fe$_2$As$_2$, \textit{Sci. Adv.} \textbf{2}, e1600807 (2016).

%\bibitem{Mizukami2014} Mizukami, Y., Konczykowski M., Kawamoto, Y., Kurata, S., Kasahara, S., Hashimoto, K., Mishra, V., Kreisel, A., Wang, Y., Hirschfeld, P.J., Matsuda, Y., Shibauchi, T. Disorder-induced topological change of the superconducting gap structure in iron pnictides, \textit{Nat. Comm.} \textbf{5}, 5657 (2014).

%\bibitem{Mizukami2014} Mizukami, Y. \textit{et.\ al}. Disorder-induced topological change of the superconducting gap structure in iron pnictides, \textit{Nat. Comm.} \textbf{5}, 5657 (2014).

%\bibitem{Hashimoto2013} Hashimoto, K., Mizukami, Y., Katsumata, R., Shishido, H., Yamashita, M., Ikeda, H., Matsuda, Y., Schlueter, J.A., Fletcher, J.D., Carrington, A., Gnida, D., Kaczorowski, D., Shibauchi, T. Anomalous superfluid density in quantum critical superconductors, \textit{Proc. Natl. Acad. Sci.} \textbf{110}, 3293 (2013).

\end{thebibliography}

\end{document}

\title{Observation of the Non-linear Meissner Effect: \\ Supplementary Information}

\author{J. A. Wilcox}
\affiliation{H. H. Wills Physics Laboratory, University of Bristol, Tyndall Avenue, Bristol, BS8 1TL, United Kingdom}

\author{M. J. Grant}
\affiliation{H. H. Wills Physics Laboratory, University of Bristol, Tyndall Avenue, Bristol, BS8 1TL, United Kingdom}

\author{L. Malone}
\affiliation{H. H. Wills Physics Laboratory, University of Bristol, Tyndall Avenue, Bristol, BS8 1TL, United Kingdom}

\author{C. Putzke}
\altaffiliation{Present address:Laboratory  of  Quantum  Materials,  Institute  of  Materials, \'Ecole  Polytechnique  F\'ed\'erale  de  Lausanne  (EPFL),  1015  Lausanne,  Switzerland}
\affiliation{H. H. Wills Physics Laboratory, University of Bristol, Tyndall Avenue, Bristol, BS8 1TL, United Kingdom}

\author{D. Kaczorowski}
\affiliation{Institute of Low Temperature and Structure Research, Polish Academy of Sciences, 50-950 Wroclaw, Poland}

\author{T. Wolf}
\affiliation{Institute for Quantum Materials and Technologies, Karlsruhe Institute of Technology, 76021 Karlsruhe, Germany}

\author{F. Hardy}
\affiliation{Institute for Quantum Materials and Technologies, Karlsruhe Institute of Technology, 76021 Karlsruhe, Germany}

\author{C. Meingast}
\affiliation{Institute for Quantum Materials and Technologies, Karlsruhe Institute of Technology, 76021 Karlsruhe, Germany}

\author{J. G. Analytis}
\altaffiliation{Present address:Department of Physics, University of California, Berkeley, California 94720, USA.}
\affiliation{Geballe Laboratory for Advanced Materials and Department of Applied Physics, Stanford University, California 94305-4045, USA}
\affiliation{Stanford Institute for Materials and Energy Sciences, SLAC National Accelerator Laboratory, 2575 Sand Hill Road, Menlo Park, California 94025, USA}

\author{J.-H. Chu}
\altaffiliation{Present address: Department of Physics, University of Washington, Seattle, WA, 98195 USA}
\affiliation{Geballe Laboratory for Advanced Materials and Department of Applied Physics, Stanford University, California 94305-4045, USA}
\affiliation{Stanford Institute for Materials and Energy Sciences, SLAC National Accelerator Laboratory, 2575 Sand Hill Road, Menlo Park, California 94025, USA}

\author{I.R. Fisher}
\affiliation{Geballe Laboratory for Advanced Materials and Department of Applied Physics, Stanford University, California 94305-4045, USA}
\affiliation{Stanford Institute for Materials and Energy Sciences, SLAC National Accelerator Laboratory, 2575 Sand Hill Road, Menlo Park, California 94025, USA}

\author{A. Carrington}
\email{Corresponding author: email a.carrington@bristol.ac.uk}
\affiliation{H. H. Wills Physics Laboratory, University of Bristol, Tyndall Avenue, Bristol, BS8 1TL, United Kingdom}

\maketitle

\subsection{Sample Dimension}

All three compounds, CeCoIn$_5$, LaFePO and KFe$_2$As$_2$ have a tetragonal structure and the single crystal samples in this study took the form of thin platelets, with the shortest dimension ($l_z$) corresponding to the $c$ axis and the larger dimensions ($l_x$ and $l_y$) corresponding to the $a$-$b$ plane. The dimensions of all samples are listed in Table \ref{table1}.

\begin{table} [h]
\centering
\begin{tabular}{ l | c | c | c }
Sample \, &\, $l_x$ ($\mu$m) \,&\, $l_y$ ($\mu$m) \,&\, $l_z$ ($\mu$m) \\
\hline
CeCoIn$_5$	& 200 & 310 & 5 \\ %C1
LaFePO	& 270 & 274 & 31 \\ %L1
KFe$_2$As$_2$	& 715 & 405 & 50 \\ %K2
\end{tabular}
\caption{\textbf{The dimensions of the samples used in this study.}}
\label{table1}
\end{table}

\subsection{Measurement geometry}
For thin, platelet samples measurements of the in-plane penetration depth can be conducted in two geometries. With the field directed along the short $c$-axis the screening currents flow only in the $ab$-plane, but the demagnetising effects will be large, with the field strongly enhanced on the $ac$ or $bc$ faces of the crystal. The field enhancement is particularly large close to the corners, and so the field will be quite non-uniform (see below for details of this).  This is clearly, not ideal for our measurements where we are concerned with the field dependence of $\lambda$.  The alternative geometry is where the field is directed along the $ab$-plane.  In this case, the demagnetisation effects are very small, and the field uniform, but the measured $\lambda_m(T)$, will be a mixture of the in-plane and the $c$-axis response.  To a good approximation, the volume penetration by field in the $H\|a$ geometry will be given by
\[
\Delta V = 2(\ell_a\ell_b \lambda_{ab} + \ell_a\ell_c \lambda_c),
\]
where $\ell_a, \ell_b,\ell_c$ are the physical dimensions along each axis.  If we then calculate an effective $\lambda$ from $\Delta V$ by dividing by the in-plane dimensions we have
\[
\lambda_e = \lambda_{ab} + \frac{\ell_c}{\ell_b}\lambda_c.
\]
As long as the samples are sufficiently thin ($\ell_c/\ell_b \ll 1$) or more precisely $(\ell_c \Delta\lambda_c(T))/(\ell_b\Delta\lambda_{ab}(T))\ll 1$ then $\lambda_e$ is very close to $\lambda_{ab}$. Demagnetising effects are avoided and the measured $\Delta \lambda(T)$ is predominately the in-plane response. 

For CeCoIn$_5$, and KFe$_2$As$_2$ the large aspect ratios led us to use the $H\|ab$ geometry for our measurements whereas for LaFePO the thicker samples and larger $\lambda_c/\lambda_a$ anisotropy meant the the $H\|c$ geometry was more suitable.

For CeCoIn$_5$, Howald \emph{et.\ al.}\cite{Howald2013} report an anisotropy in the zero temperature penetration depth $\gamma = \lambda_c / \lambda_a \simeq 1.3$ and over the temperature range $T=0$ to $T/T_c = 0.25$, the ratio $\Delta\lambda_c / \Delta\lambda_a \simeq 4.4$.  Our sample has an aspect ratio of approximately 50 (Table S1), which means that the $c$-axis contribution to $\lambda_e$ would be $\sim 10$\%.  However, Howald \emph{et.\ al.} find a much weaker temperature dependence of $\Delta\lambda_a$ compare to our current data, possibly because of sample purity issues. If we use our own data for $\Delta \lambda_{ab}(T)$ in conjunction with $\Delta \lambda_c (T)$ from Howald \emph{et.\ al.}\ the $c$-axis contribution drops to around 2\%. In both cases this is a relatively small correction.

\begin{figure}
\centering
\includegraphics[width=0.80\linewidth]{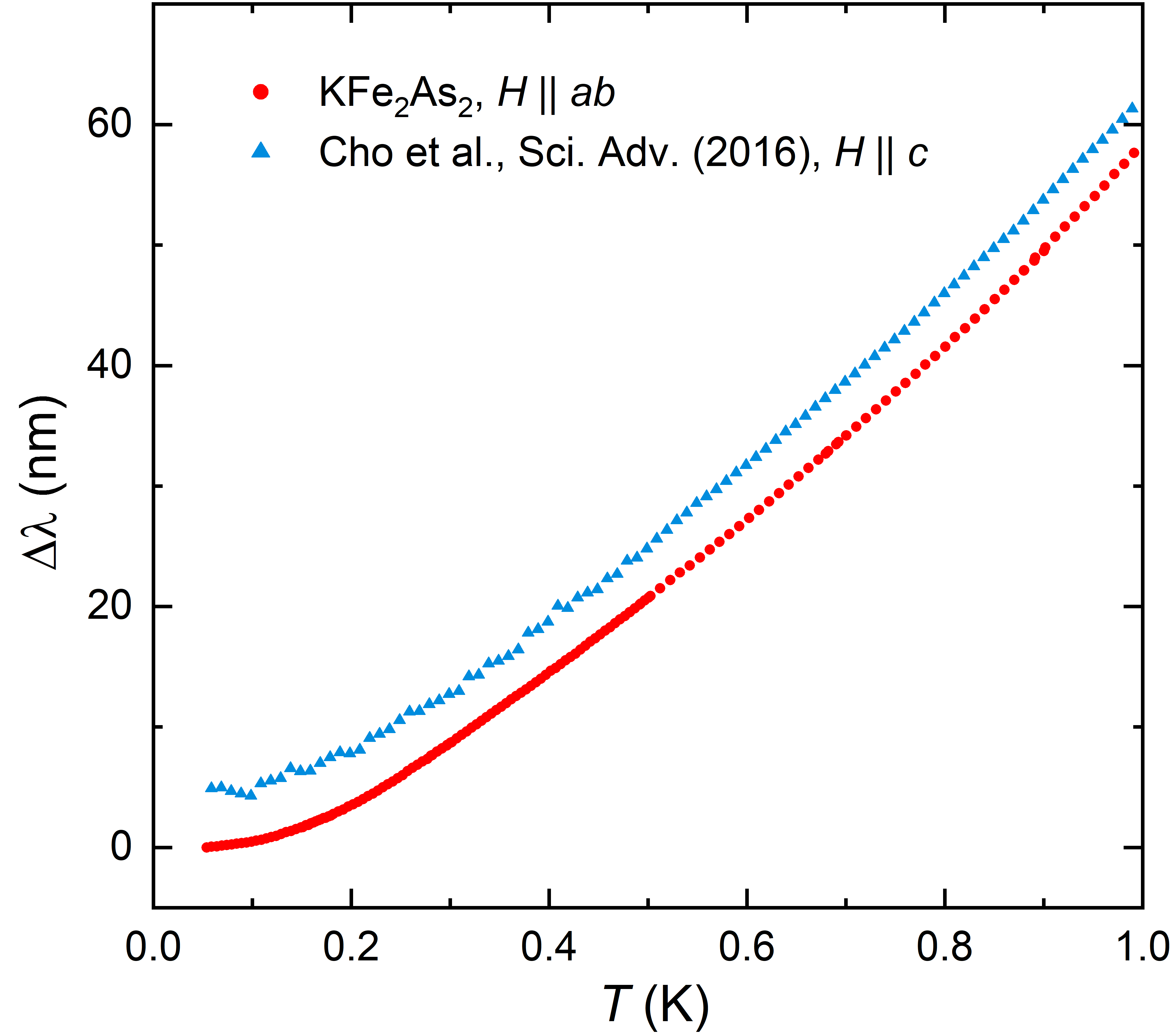}
\caption{\textbf{Comparison between measurements of $\Delta\lambda(T)$ for KFe$_2$As$_2$.} Our data for KFe$2$As$_2$, measured in the $H\|ab$ orientation is compared to the in-plane response reported in Ref.\ \cite{Cho2016} (offset by 5nm along the $\Delta\lambda$ axis for clarity).}
\label{KFA_comparison}
\end{figure}

In the case of KFe$_2$As$_2$, we are not aware of direct measurements of $\lambda_c(T)$, however, the anisotropy in the absolute values at $T=0$ has been measured to be relatively weak, with $\gamma=3.6$ \cite{Kawano2013}. The response for our KFe$_2$As$_2$ sample, in the $H\|ab$ orientation, is shown in direct comparison to the in-plane response $\Delta\lambda_{ab}(T)$ from Ref.\ \cite{Cho2016}. The agreement between the two measurements is very good and shows that there is very little $c$-axis contribution to the response over the temperature range in question.

\subsection{Determination of Field of First Flux Penetration}

\begin{figure}
\centering
\includegraphics[width=0.9\linewidth]{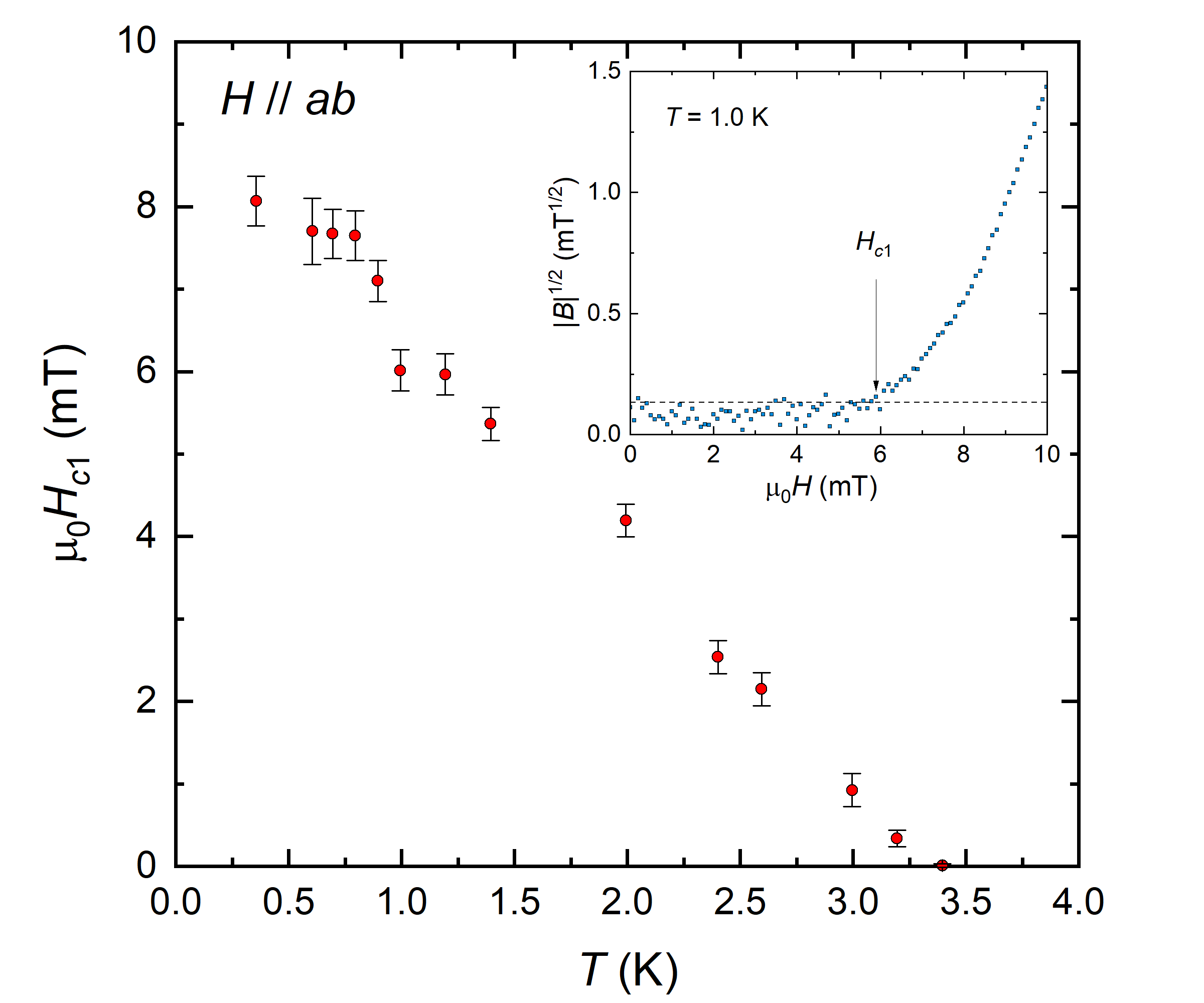}
\caption{\textbf{Temperature dependence of the lower critical field $H_{c1}$ of KFe$_2$As$_2$.} For these measurements the magnetic field is oriented in-plane ($H\|ab$). The inset shows an example response of a Hall sensor located beneath the edge of the sample. $H_{c1}$ is the field at which the measured flux density begins to increase.}
\label{KFA_Hc1}
\end{figure}

\begin{figure}
\centering
\includegraphics[width=0.9\linewidth]{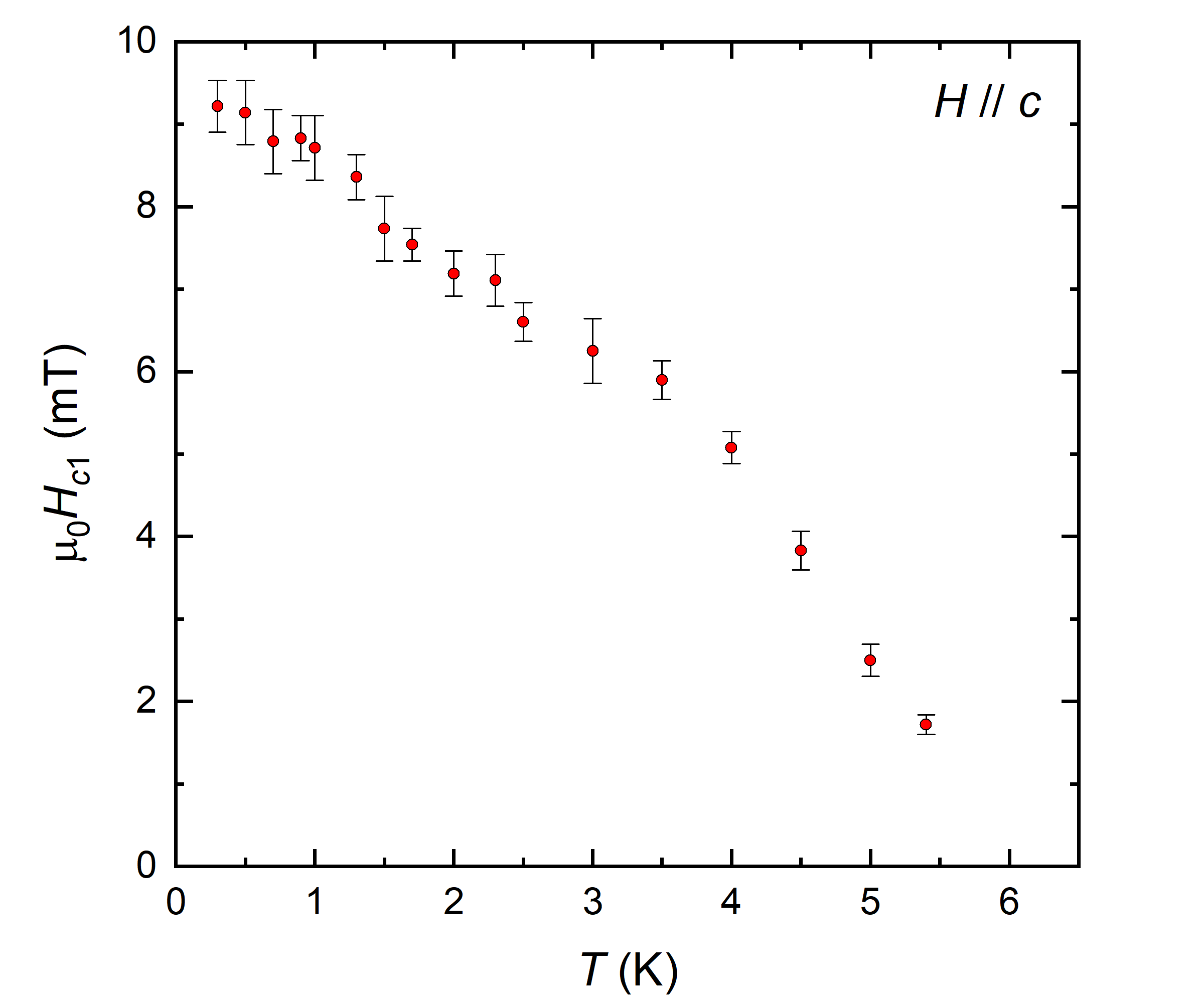}
\caption{\textbf{Temperature dependence of the lower critical field $H_{c1}$ of LaFePO.} In these measurements the magnetic field is oriented parallel to the $c$-axis ($H\|c$).}
\label{LFPO_Hc1}
\end{figure}

For KFe$_2$As$_2$ and LaFePO, the lower critical field $H_{c1}$, or more precisely the field of first flux penetration $H_p$, was determined in the same samples used for the $\lambda(T,H)$ study.  The samples were suspended above a micro-Hall probe array, cooled to base temperature and then a dc field was applied as described in Ref.\ \cite{Putzke2014}. For KFe$_2$As$_2$ $H$ was parallel to the $ab$-plane and for LaFePO $H\|c$ as in the $\lambda(T,H)$ study. At low field the local field measured by the Hall probe is linear in $H$ corresponding to the field leaking around the sides of the sample.  At $H_p$ there is a sharp rise in $B$ as seen in the inset to Fig.\ \ref{KFA_Hc1}.  This field, which is equal to $H_{c1}$ in the absence of any surface barriers is plotted in Figs.\ \ref{KFA_Hc1} and \ref{LFPO_Hc1}.  $H_{c1}(T)$ is found to have a linear temperature dependence, mirroring that found for $1/\lambda(T)^2$. For LaFePO the field has been corrected for demagnetising factors using the method of Brandt \cite{Brandt1999,Putzke2014}.  As the fields are small, sweeps were made in positive and negative fields to correct for the Earth's field.

%are there other measurements of HC1 for KFe2As2

\begin{figure}
\centering
\includegraphics[width=0.9\linewidth]{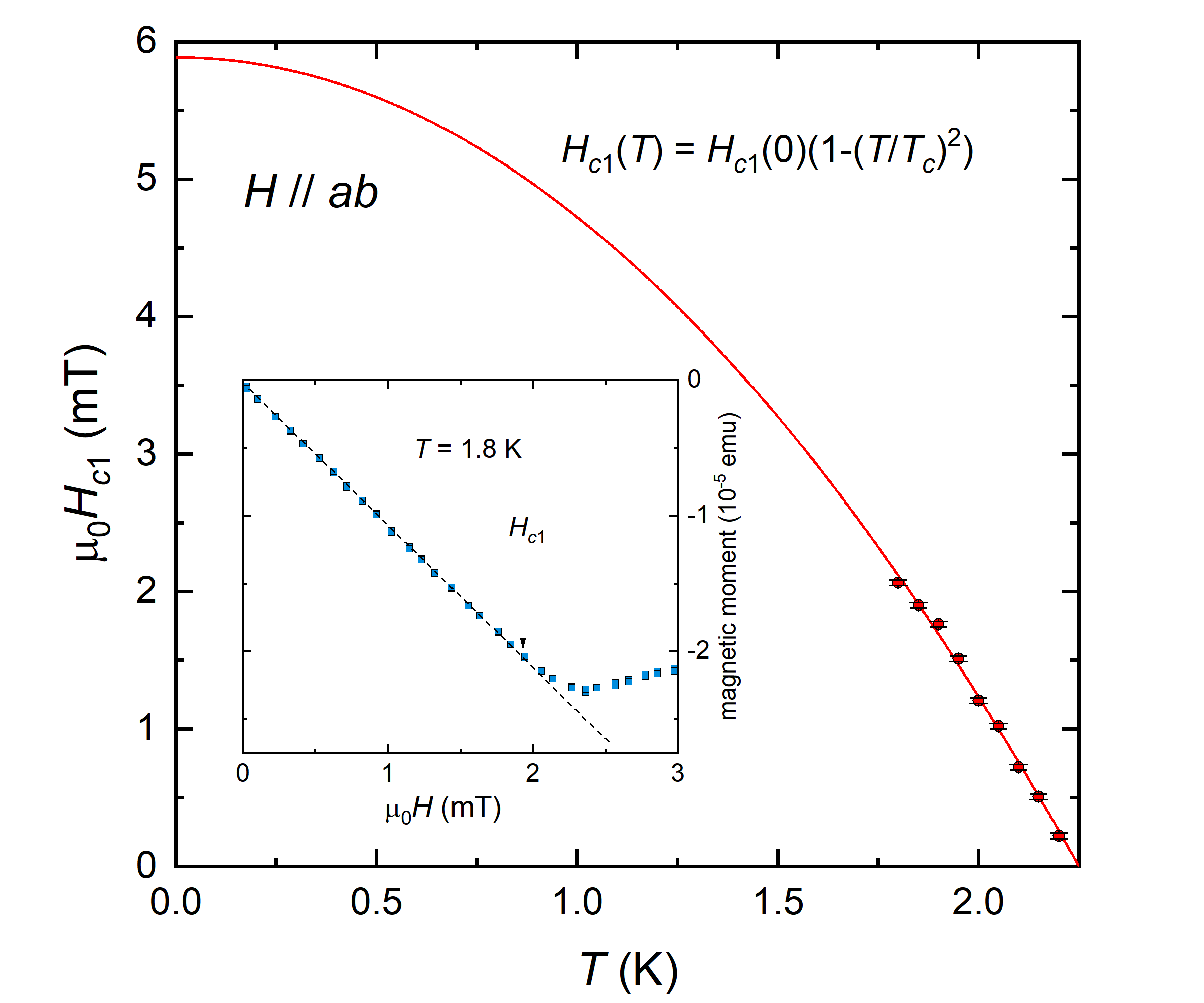}
\caption{\textbf{Temperature dependence of the lower critical field $H_{c1}$ of CeCoIn$_5$.} For these measurements the magnetic field is oriented in-plane ($H\|ab$) and the values of $H_{c1}$ are determined from bulk magnetisation using a SQUID magnetometer. The inset shows an example measurement for positive field, where $H_{c1}$ is taken as the field at which the susceptibility $\chi$ deviates from $-1$ (as indicated by the dashed line).}
\label{CCI_Hc1}
\end{figure}

For CeCoIn$_5$, a different method was used because of the unavailability of apparatus. Instead $H_{c1}$ was measured using a commercial SQUID magnetometer. The magnetic moment was measured as a function of field and the point where $m(H)$ departs from linearity identified as $H_{c1}$ (see insert to Fig.\ \ref{CCI_Hc1}).  The data only extend down to $T=1.7$\,K and so were extrapolated to lower temperature using the standard empirical formula $H_{c1}(T)=H_{c1}(0)(1-(T/T_c)^2)$.  Although this extrapolation is only appropriate for conventional fully gapped superconductors, a linear extrapolation as suggested by the linear $T$ dependence of $1/\lambda^2(T)$ would give a higher value of $H_{c1}(0)$, so the extrapolation in Fig.\ \ref{CCI_Hc1} ($H_{c1}(0) =5.9 \pm 0.1$\,mT) should be viewed as a lower limit. The results presented are similar to those found in the literature; performing the same analysis on the values of $H_{c1}(T)$ reported by Majumdar \emph{et.\ al.}\cite{Majumdar2003} yields a result of $H_{c1}(0) =4.52 \pm 0.02$\,mT.

\subsection{Paramagnetic impurities and Andreev bound states}

Impurities, both magnetic and non-magnetic can affect the measured behaviour of $\lambda(T,H)$ as can surface Andreev bound states.  In this section we consider the effect of paramagnetic impurities and Andreev bound-states.  Non-magnetic impurities are considered in the next section.

\textbf{Paramagnetic impurities} can contribute to the measured penetration depth, giving a contribution which is proportional to the normal state susceptibility $\chi_N(T)$ \cite{Cooper1996,Serafin2010}. A flattening or an upturn at low temperatures in $\lambda(T)$ can then result from $\chi_N(T)$ following a Curie-law behaviour if the concentration of paramagnetic ions is sufficiently large.  The effect of finite field on these paramagnetic terms can be estimated from the calculated field dependence of $\chi_N(T)$.

For classical paramagnetism, the induced magnetisation $M$ in a magnetic field $B$ is given by
\begin{equation}
M = N g_J \mu_\mathrm{B} J F_J(x) ,
\label{Eq:MBF}
\end{equation}
where $N$ is the number of moments per unit volume, $g_J$ is the Land\'{e} $g$-factor, $J$ is the total electronic angular momentum and $F_J(x)$ is the Brillouin function,
\begin{equation}
F_J(x) = \frac{2J+1}{2J}\coth\left(\frac{2J+1}{2J}x\right)-\frac{1}{2J}\coth\left(\frac{1}{2J}x\right) ,
\label{Eq:BF}
\end{equation}
in which $x$ is defined as
\begin{equation}
x = \frac{J g_J \mu_\mathrm{B} B}{k_\mathrm{B} T}.
\label{Eq:xBF}
\end{equation}
From this we can calculate the susceptibility 
\begin{equation}
\chi = \frac{dM}{dH} =\frac{N \mu_0(g_J \mu_\mathrm{B} J)^2}{k_\mathrm{B} T} \frac{dF_J(x)}{dB}
\label{Eq:susceptibilityBF}
\end{equation}
The susceptibility of these moments will then contribute to the measured penetration depth $\lambda_m(T)$ according to
\begin{equation}
\lambda_m(T) \simeq \lambda_\mathrm{L}(T)\sqrt{1 + \chi(T)} ,
\label{Eq:PMlambda}
\end{equation}
where $\lambda_\mathrm{L}(T)$ is the London penetration depth (i.e.\ no impurities) \cite{Cooper1996}. For small amounts of impurities, $\chi \ll 1$, the change in the penetration depth with temperature can be expressed as \cite{Serafin2010}
\begin{equation}
\Delta\lambda (T)= \Delta\lambda_L(T)  + \chi(T) \lambda_0/2 .
\label{Eq:parmaglambda}
\end{equation}

\begin{figure}[h]
\centering
\includegraphics[width=0.9\linewidth]{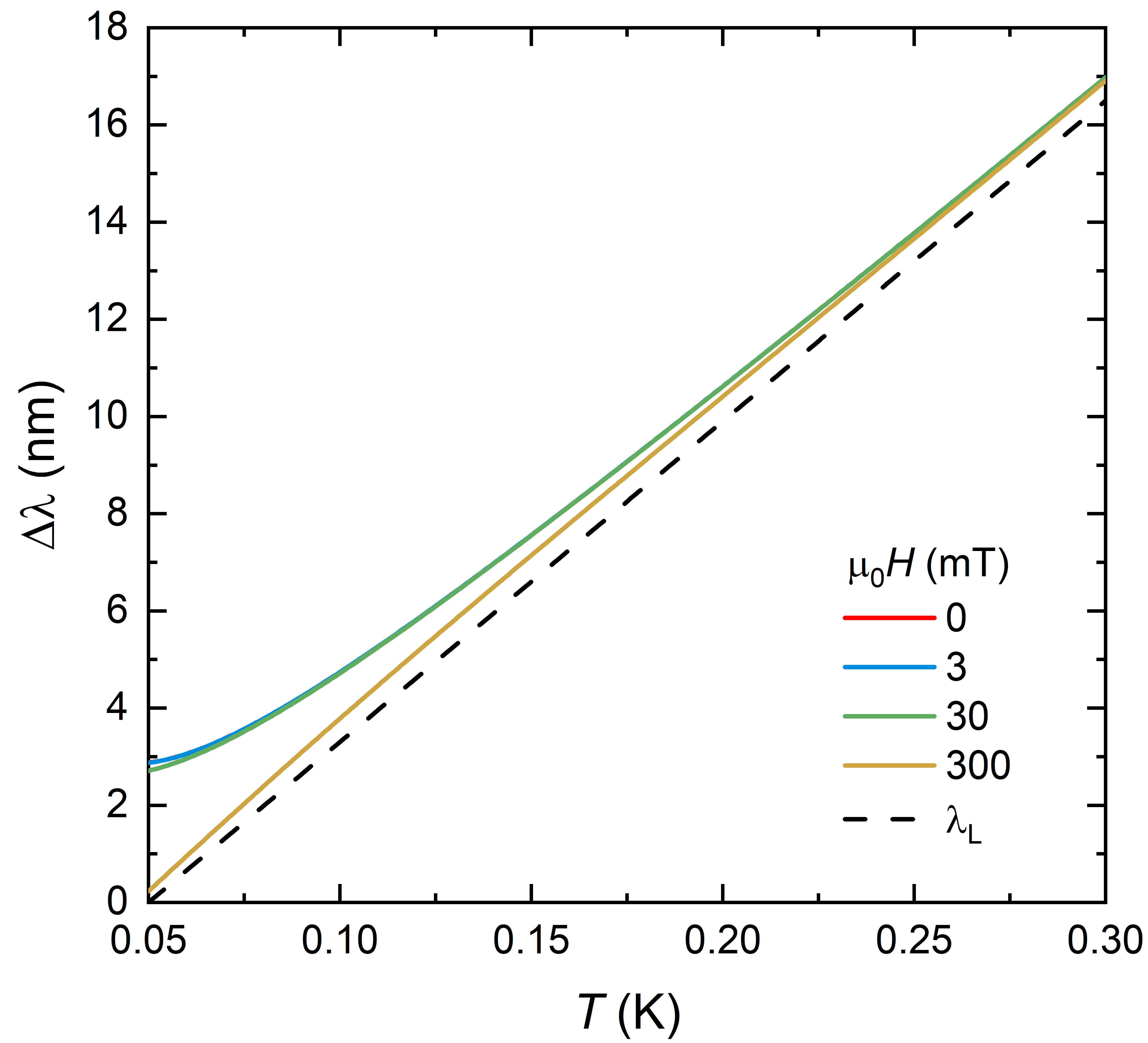}
\caption{\textbf{Calculation of the expected change in the magnetic penetration depth due to paramagnetic impurities} The response of Ce$^{3+}$ impurities in fields of 0 (red curve), 3 mT (blue curve), 30 mT (green curve) and 300 mT (yellow curve), compared to a pure linear response (dashed black line).} %needs to be mu_0 H in figure
\label{paramag_calc}
\end{figure}

As an example, we consider the case of Ce$^{3+}$ ($S=1/2, J=5/2$), which is a plausible impurity in CeCoIn$_5$. We assume a pure linear dependence of $\Delta\lambda_\mathrm{L}(T)$ and add a small paramagnetic contribution which, in the limit of zero applied field, gives a characteristic flattening of $\Delta\lambda(T)$ at low-$T$ (Fig.\ \ref{paramag_calc}).  Increasing the field to 3\ mT (maximum experimental value for  CeCoIn$_5$), gives a result almost indistinguishable from the zero field curve.  A further increase to  $\mu_0H = 30$ mT gives a very small decrease in $\Delta\lambda$ of approximately 0.2\ nm at 50\ mK. Only with a field 100 times larger than used experimentally (300\ mT) is a significant response induced, and in that case the effect is to decrease $\Delta\lambda$, i.e., the change is in the opposite direction to that observed for CeCoIn$_5$.  This large difference in field scale between the effect of finite field on paramagnetic impurities compared to that for the non-linear-Meissner effect is the key point which allow us to distinguish between the two effects.   Note that for lower spin species (e.g., $J=1/2$) the effect would be even weaker.

\textbf{Andreev bound states (ABS)} can give an additional contribution to the low temperature penetration depth. In the case of the cuprates, where there is a single sign-changing gap due to the $d_{x^2-y^2}$ pairing state, zero energy ABS occur on sample surfaces which are not perpendicular to the principal crystallographic directions; (100), (010), or (001) \cite{Hu1994,Fogelstrom1997}.  These zero energy states are broadened in energy at finite temperature and contribute a zero energy peak to the density of states and hence give rise to an upturn in $\lambda(T)$ at low temperatures if the field is probing ABS surfaces \cite{Barash2000,Walter1998,Carrington2001}.  This upturn is suppressed in small fields because the ABS split and shift higher in energy with increasing field \cite{Barash2000,Carrington2001}. This increase in temperature dependence of $\lambda(T)$ with increasing field is opposite to what we find for CeCoIn$_5$ or LaFePO and hence any ABS contribution must be small compared to the intrinsic bulk non-linear response.  This is consistent with the $H\|ab$ measurement geometry used for CeCoIn$_5$. With this geometry the main contribution to $\Delta\lambda(T)$ comes from the (001) faces on which ABS cannot form.  ABS contribution were not detected for YBa$_2$Cu$_3$O$_{6+x}$  in this geometry for the same reason \cite{Carrington2001}.

For iron-based superconductors with a sign-changing gap, theory suggests that any ABS would occur at finite energy  \cite{Araujo2009,Huang2010} and so would not produce a zero-energy peak in the DOS and hence would not give rise to an low-$T$ upturn in $\lambda(T)$ or the associated field dependence.  For LaFePO, the measurement geometry $H\|c$ would in-principle allow for ABS contributions as the sample has both (100) and (110) facets \cite{Fletcher2009}, however, the observed increase in $d\lambda/dT$ with increasing field would suggest that any contribution was either small or entirely absent.  For KFe$_2$As$_2$ the measurement geometry, $H\|ab$,  means predominately (001) surfaces are probed and this together with the above mentioned theoretical work suggests that there are also no ABS contributions.

\subsection{Calculations of field dependence of $\lambda(T)$}

For our calculations of the field dependence of $\lambda(T)$ we consider a two-dimensional, circular Fermi surface with isotropic Fermi velocity $v_F$.  The gap structure is taken as either the classic $d$-wave form
\begin{equation}
  \Delta(\phi,T)=\Delta_0(T) \cos(2 \phi)
  \label{Eq_dgap}
\end{equation}
or a similar form but with a finite gap
\begin{equation}
  \Delta(\phi,T)=\Delta_0(T) (|\cos(2 \phi)| + \eta).
    \label{Eq_dpsgap}
\end{equation}
Note that $\lambda(T,H)$ is only sensitive to the absolute magnitude of $\Delta(\phi)$, not its sign.  The second, finite gap form could represent a mixed order parameter $d+is$ but is also representative of other strongly anisotropic forms where $\Delta$ does not change sign.  $\phi$ is the azimuthal angle.   The temperature dependence of the gap is taken to be the weak-coupling $d$-wave form in both cases, which is approximated by
\begin{equation}
    \Delta_0(t) = \Delta_0(0) \tanh\left[\frac{\pi}{\Delta_0(0)}\sqrt{\frac{4}{3}\left(\frac{1}{t}-1\right)}\right]
\end{equation}
where $\Delta_0(0)=2.14$ and $t = T/T_c$.

To calculate $\lambda(t,H)$ we used the following expression \cite{Stojkovic1995,Xu1995} for the non-linear quasiparticle current $ j_{qp}$ in the $x$-direction ($\phi=0$) created by the applied field, integrated over the Fermi surface
\begin{equation}
    j_{qp} = A \int_{-\pi}^{\pi} d\phi \mathcal{R} \sum_n  \frac{(\sigma(\phi)-i\omega_n) \cos(\phi)}{\sqrt{(\omega_n+i\sigma(\phi,t))^2+|\Delta(\phi)|^2}}
    \label{qpcurrent}
\end{equation}
where $\omega_n = (2n+1)\pi t$ are the Matsubara frequencies,
\begin{equation}
\sigma(\phi)= \alpha H \cos(\phi-\phi_0)
\end{equation}
is the quasiparticle energy shift produced by the field $H$ induced superflow and $\mathcal{R}$ denotes the real part of the sum. The limits of the sum over $n$ are $\pm\infty$ but practically we set a maximum frequency $|\omega_0| = 100$ (in units of $T_c$).  Convergence was checked by varying $|\omega_0|$ up to 2 orders of magnitude higher and checking the $H$ and $T$ dependent results were insensitive.  The direction of the field is set by $\phi_0$. $A$ is a constant related to the normal state parameters which determine the absolute value of $\lambda(0)$, and $\alpha =\mu_0 v_F \lambda(0) e/(k_B T_c)$ is the constant of proportionality between the field and energy shift (normalised to $k_B T_c$).   

The temperature and field dependence of the normalised superfluid density is then calculated from
\begin{equation}
\frac{\lambda^{2}(0)}{\lambda^{2}(t,H)} = C\frac{j_{qp}}{H},
\end{equation}
where $C$ is a constant determined by ensuring $\lambda^{2}(0)/\lambda^{2}(t,H)=1$ as $(t,H)\rightarrow 0$.
This represents an approximation which is exact in the linear-response limit, and gives the correct slope d$\log\lambda/$d$t=\ln 2 /(\Delta_0/T_c)$ as $(t,H)\rightarrow 0$ \cite{Xu1995} for the $d$-wave gap (Eq.\ \ref{Eq_dgap}).  To get the exact result for finite $H$, the non-linear London equations should be solved, combining the result for $q_{qp}$ with Maxwell's equations, in the appropriate geometry. However, it was shown in Refs.\ \cite{Stojkovic1995,Xu1995} that in the usual semi-infinite plane geometry this only changes the field scale by a factor $3/2$ without otherwise changing the field or temperature dependence of $\lambda$.  Our approach mirrors that of Xu \textit{et al.} \cite{Xu1995} who solved the non-linear London equations analytically for the pure, zero temperature field dependence of $\lambda$, but for finite temperature/scattering they calculated numerically the current-field relation and scaled this to the pure, $T=0$ result to get the non-linear response of $\lambda$.  Our finite $H$ results for $\lambda(H)$ are consistent with those of Stojkovi\'c and Valls \cite{Stojkovic1995} who did solve the non-linear London equations at finite $T$.

\begin{figure}[b]
\centering
\includegraphics[width=0.9\linewidth]{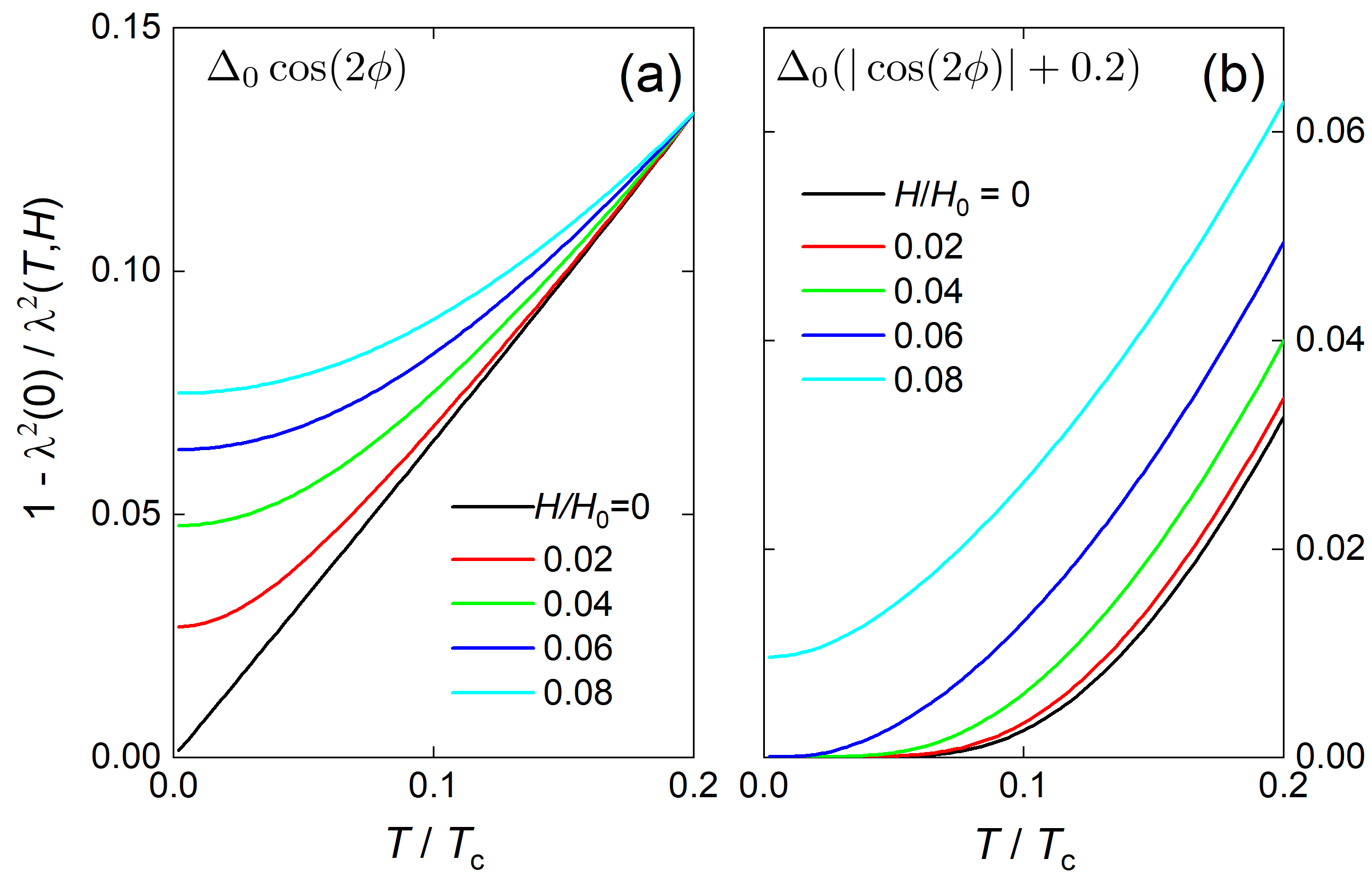}
\caption{\textbf{Calculated temperature dependence of the normalized superfluid density $\lambda^2(0)/\lambda^2(t,H)$.} The fields used in the calculations are indicated (in units of $H_0$) and the calculations are in the pure limit (no impurities). (a) $d$-wave gap (Eq.\ \ref{Eq_dgap}) (b) ($d+is$)-gap (Eq.\ \ref{Eq_dpsgap} with $\eta=0.2$).}
\label{calclamTB}
\end{figure}

The results for the two different gap forms, $d$-wave gap (Eq.\ \ref{Eq_dgap}) and ($d+is$)-gap [Eq.\ \ref{Eq_dpsgap} with $\eta=0.2$] in the pure limit ($\Gamma=0$) are shown in Fig.\ \ref{calclamTB} (a) and (b).  The magnetic fields in the calculations are scaled according to
\begin{equation}
    \mu_0H_0 = \frac{3 \Delta_0(0)}{e\lambda(0)v_F}
\end{equation}
which takes into account the above factor $3/2$.   In the main text we subtract a constant from each curve at finite $H$ so that $\Delta\lambda(t)$ coincides with the $H=0$ result at $t=0.2$ for easy comparison with the experimental data.

\begin{figure}
\centering
\includegraphics[width=0.9\linewidth]{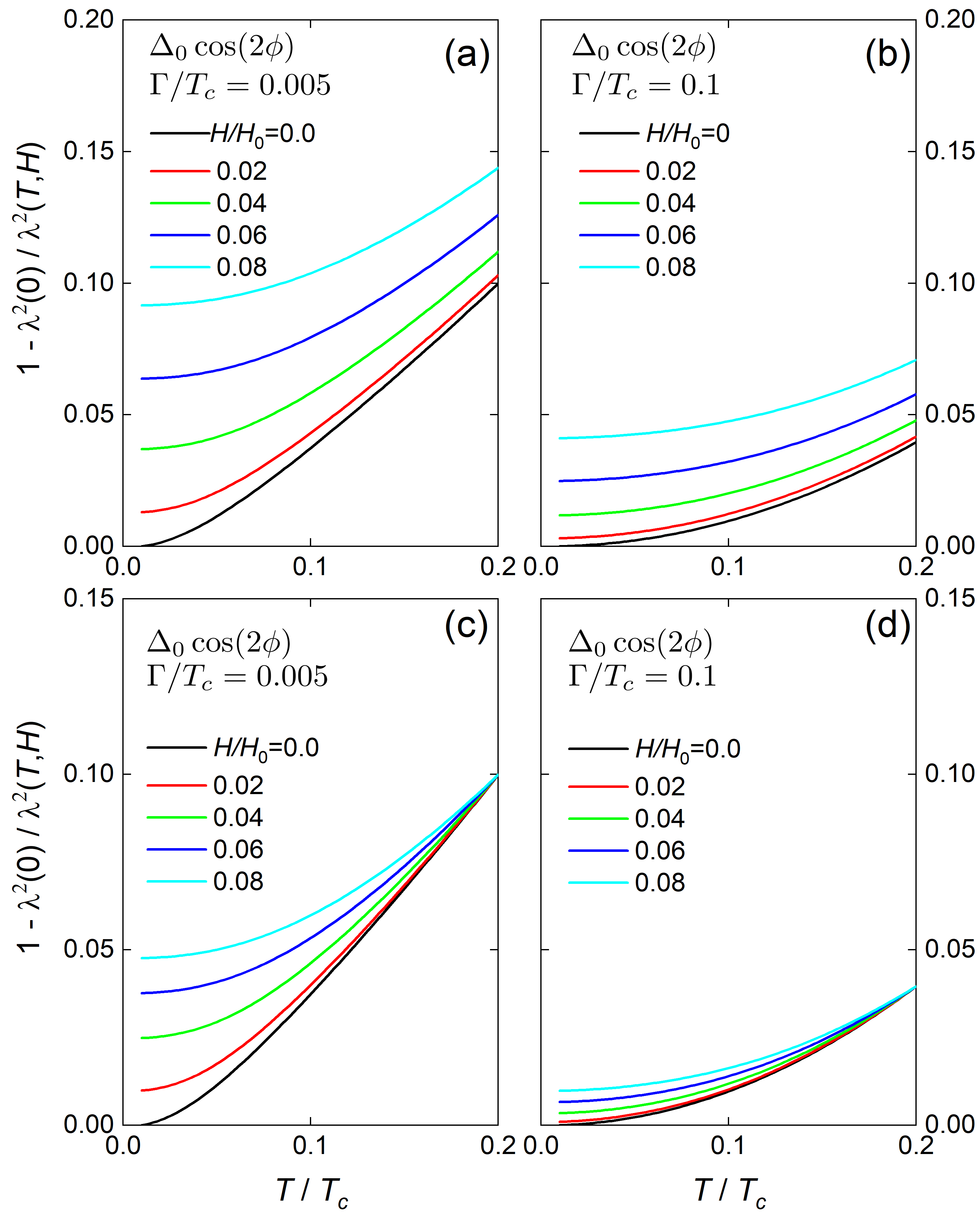}
\caption{\textbf{Effect of impurities on the non-linear response for a $d$-wave gap} (a) low impurity density,  $\Gamma=0.005$, (b) high impurity density, $\Gamma=0.1$. (c) and (d) are the same data shifted  vertically down so that they coincide with the $H = 0$ results at $T/T_c = 0.2$ as in figure 1 of the main text.}
\label{calclamTBimp}
\end{figure}
\vskip 12pt
\noindent
\textbf{Non-magnetic impurities} are included by replacing the Matsubara frequencies $\omega_n$ with their renormalised values
\begin{equation}
\tilde{\omega}_n = \omega_n + \Gamma \frac{N(\tilde{\omega}_n,t)}{c^2 + N^2(\tilde{\omega}_n,t)}
\label{EqRenormMatsubara}
\end{equation}
where 
\begin{equation}
N(\omega,t) = \frac{1}{\pi}\int_{-\pi/2}^{\pi/2} \frac{\omega d\phi}{\sqrt{\omega^2+\Delta^2(\phi,t)}}
\label{EqQOS}
\end{equation}
is the quasiparticle density of states, $c$ is the cotangent of the scattering phase shift and $\Gamma = (1+c^2)\Gamma_N$, where $\Gamma_N$ is the normal state scattering rate \cite{Xu1995}.   Equations (\ref{EqRenormMatsubara}) and (\ref{EqQOS}) are solved iteratively until convergence is reached for each Matsubara frequency, \textit{i.e.}, $N(\omega_n,t)$ is first calculated with the unrenormalised $\omega_n$, then $\tilde{\omega}_n$ is calculated from Eq.\ (\ref{EqRenormMatsubara}) and this is used to calculate $N(\tilde{\omega}_n,t)$. This is repeated until $\tilde{\omega}_n$ has converged.  The reduction of the superfluid density with increasing $\Gamma_N$ calculated by this procedure as $(t,H)\rightarrow 0$ is in excellent agreement with the results of Deepwell \textit{et al.} \cite{Deepwell2013} for both the unitary and Born limits.  Our calculations of $\lambda(T,H,\Gamma)$ are in good agreement with the results of Refs.\ \cite{Xu1995} for the unitary limit ($c=0$) case considered there.   

Results are shown in Fig.\ \ref{calclamTBimp} for the $d$-wave gap function and two different values of $\Gamma$ to illustrate small and large amounts of impurity scattering in the unitary limit respectively.  It can be seen that for a small impurity density, similar to that which would cause a flattening of $\Delta\lambda(T)$ at very low temperature only, the effect on the non-linear response is quite small.  The normalised $\Delta\lambda(T)$ still increases with increasing field.  Only when the impurity scattering is large, so that  $\lambda(T)\sim T^2$ over the whole low temperature range is the non-linear response substantially reduced.

\subsection{Exponent analysis}

Whether the shifted $\Delta\lambda(T,H)$ (as in main text Fig.\ 1) increases or decreases with $H$ at the lowest temperatures for the case of a finite gap ($\eta>0$) depends on the size of the field with respect to the size of the finite gap and also the temperature at which the data are shifted to coincide. An alternative way to view the results is to fit the data to a power-law as in Fig.\ 5 of the main text.  In the case that the data do not follow a power-law over the full temperature range of the fit, the derived exponent will be a temperature-averaged value.    

\begin{figure}
\centering
\includegraphics[width=\linewidth]{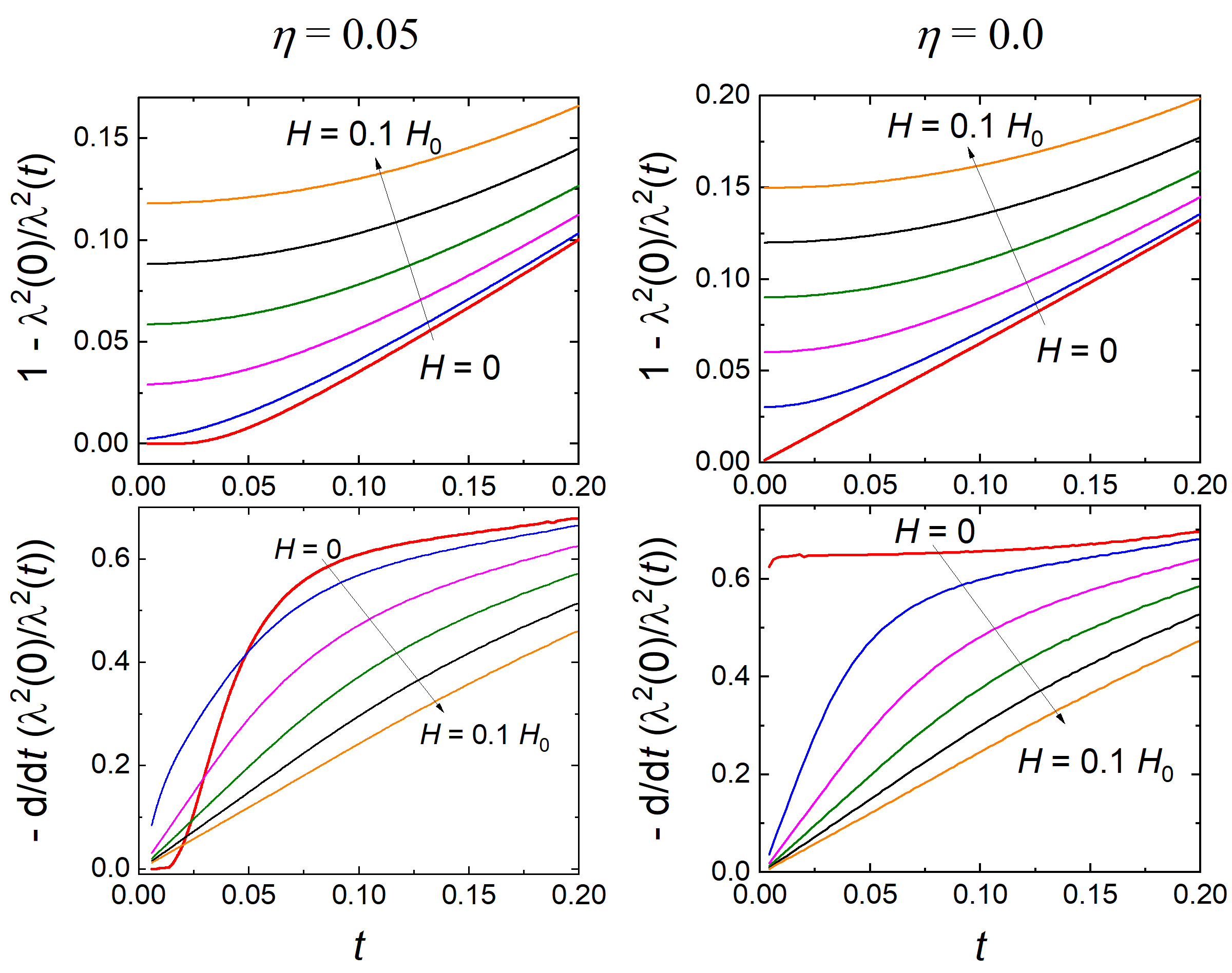}
\caption{\textbf{Exponent analysis of the calculated superfluid density}. Left panels for the small gap case ($\eta=0.05)$ and right panels for the pure-d-wave case ($\eta=0)$. Top panels: $\lambda^2(0)/\lambda^2(t,H)$ for the fields indicated. Bottom panels: the derivative  $\rm{d}\lambda^2(0)/\lambda^2(t,H)/\rm{d}t$.}
\label{Fig:ExpAnalysis_etacombined}
\end{figure}

A second alternative is to calculate the local temperature exponent, from  $n=\rm{d}\ln(\lambda(T)-\lambda(0))/\rm{d}\ln T$.  The result of doing this for the calculation with a very small gap, $\eta=0.05$, is shown in Fig.\ \ref{Fig:ExpAnalysis_etacombined}.  For zero field,  $n$ at low temperature tends to a high value indicative of an exponential $T$ dependence of $\lambda$ as expected from the model. For $H/H_0=0.02$, $n$ has decreased to $\sim 1.5$ at $t=0.03$, consistent with a much stronger (almost linear) $T$ dependence of $\lambda$.  For larger fields, $n$ tends to 2 at the lowest temperatures which is the same as expected for the nodal case with impurities.  The derivative d$\lambda^{-2}$/d$T$ shows consistent behaviour, increasing from zero for zero field (exponential behavior) to a maximum for $H/H_0\simeq 0.02$ and then reduces monotonically for higher fields.  For this gapped case, the temperature dependence at the lowest temperature is always enhanced by field at the lowest temperatures with the maximum enhancement occurring for a particular range of small applied field.   For the nodal case ($\eta=0$) (right panels Fig.\ \ref{Fig:ExpAnalysis_etacombined}), the low temperature exponent is close to 1 at zero field, and then increases monotonically to 2 at higher fields.   Unfortunately, the noise level is too great to perform a similar analysis on the experimental data, so we use the power-law fitting over an extended temperature range method in the main text.

\subsection{Demagnetisation effects}

\begin{figure}
\centering
\includegraphics[width=0.9\linewidth]{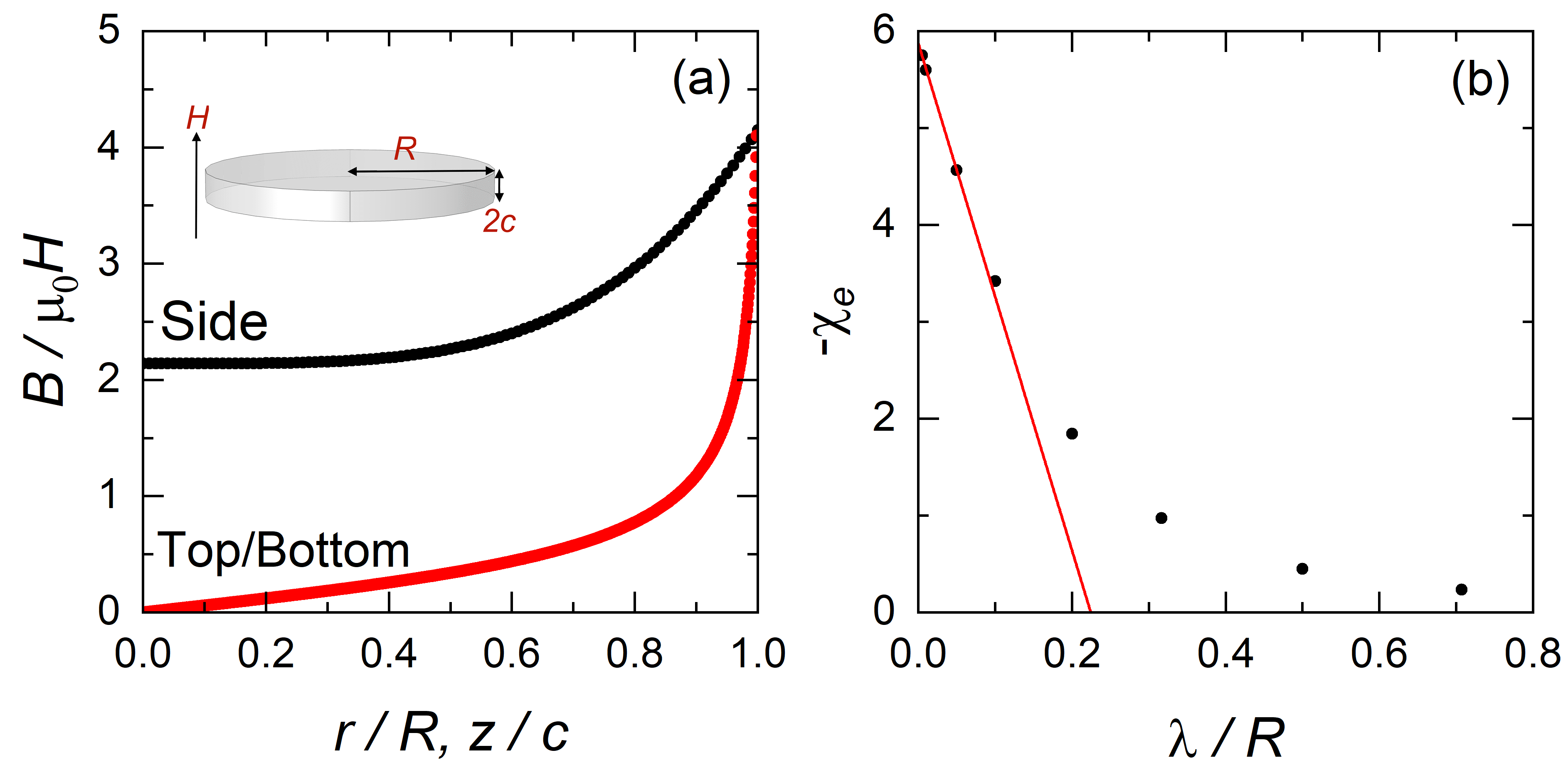}
\caption{\textbf{Effect of demagnetisation on response}. (a) The calculated variation of the tangential surface fields along the sides and top/bottom faces. The inset shows the geometry of the disc sample for the simulation. (b) The calculated change in the effective susceptibility $\chi_e$ as a function of $\lambda$.} 
\label{Fig:Demag}
\end{figure}

As discussed above, for a platelet sample in the $H\|c$ geometry, demagnetisation effects cause the surface fields to be non-uniform.  To estimate the effect of this on the measured field dependence of $\lambda$ we have performed finite-element modelling to determine the surface fields.  We chose a simplified geometry of a disk with large aspect ratio ($R:c$ = 10:1) similar to that of our LaFePO sample (see Table \ref{table1} and Fig.\ \ref{Fig:Demag}(a)).  We use the COMSOL package to solve the \textit{linear} London equation inside the disc and Ampere's outside. We solve for the vector potential $\bm{A}$ subject to the boundary condition that at large distances the $B$ field is uniform and directed along the $z$-direction.

The variation of the fields along the side and top/bottom faces are shown in Fig.\ \ref{Fig:Demag}(a). Along the sides, at the centre the field is enhanced by $\sim$ 2 and this increases as we move towards the top edge.  Along the top edge, moving from the edge towards the centre, the field decreases rapidly initially and then goes approximately linearly to zero at the centre of the top face.  Integration of the current $\bm{j}$ distribution inside the sample volume ($V$) gives the moment, $m$ from which the effective susceptibility $\chi_e$ can be calculated: 
\[
m=\frac{1}{2}\int (\bm{r}\times \bm{j})dV = -\mu_0 \lambda^{-2} \int (\bm{r}\times \bm{A})dV ,
\]
and
\[
\chi_e = \frac{m}{V H},
\]
where we have used the linear London equation $\bm{j} =-\bm{A}/(\mu_0 \lambda^2)$.

In Fig.\ \ref{Fig:Demag}(b) we show the results of performing this calculation for different values of $\lambda$. Experimentally, it is useful to calculate the ratio of frequency changes in our resonant coil as $\lambda$ varies to the change when the sample is completely removed from coil. In the limit that $R\gg \lambda$ this ratio is a linear function of $\lambda$ and can be expressed as an effective dimension of the sample,  $\Delta F_\lambda/\Delta F_0 = \lambda /R_{3D}$.  The calculation for our disc gives $R_{3D}=0.22 R$ which compares well to the value estimated in Ref.\ \cite{Prozorov2006} where in the thin limit $R\gg c$, $R_{3D}=1/(2+\pi)R=0.194R$.

To calculate the non-linear response would require solving the non-linear London equations for our geometry which is outside of the scope of the current work. We can however estimate the effect of the field variation across the sample by calculating the volume penetrated by the field into the sample.  If we assume a linear response, the penetration is purely exponential and as $\int_0^\infty \exp(-x/\lambda)=\lambda$, we can estimate the volume penetration $V_e$ into the sample from the surface fields, by integrating over the top/bottom and side surfaces.   For our disc,
\begin{equation}
V_e= \int_{0}^c  4\pi R \frac{H(z)}{H_a} \lambda dz + \int_0^R 4\pi r \frac{H(r)}{H_a} \lambda dr.
\label{Eq:volpen}
\end{equation}

$V_e$ underestimates the difference $m(\lambda)-m(\lambda=0)$, compared to $\chi_e$ by approximately a factor 2 for our disc, because it neglects the changes in the fields outside of the sample. However, it is reasonable to assume we can nevertheless use $V_e$ to estimate the non-linear response by setting  $\lambda$ in Eq.\ \ref{Eq:volpen} to $\lambda(H)=\lambda(0)(1+\alpha H)$, and then calculate the field dependent enhancement of $\lambda$ from $V_e(\lambda(H))/V_e(\lambda(0))$.  Setting this ratio equal to $1+\alpha^\prime$, we find that $\alpha^\prime / \alpha = 1.09$, so that the response for this disk is almost the same as the local response (i.e., the response without demagnetising effects).  The reason for this is that the effect of the field enhancement on some parts of the sample is cancelled out by the parts close to the centre of the top/bottom faces where the field is lower.  

\subsection{Estimation of size of $\Delta\lambda(B)$ for LaFePO and CeCoIn$_5$}
In general, for a single band $d$-wave superconductor, $H_0$ is given by \cite{Xu1995}
\begin{equation}
    \mu_0H_0 = \frac{3 \mu_\Delta \Delta_0}{2e\lambda_0v_F}
    \label{Eq:H0}
\end{equation}
where now, the angular slope of the gap at the node $\mu_\Delta=\frac{1}{\Delta_0}\frac{d\Delta}{d\phi}_{\rm node}$ is allowed to vary ($\mu_\Delta=2$ for the $d$-wave model in Eq.\ \ref{Eq_dgap}). 

A fit to the linear section of $\Delta\lambda(B)$ (Fig.\ 3, main text) to
\begin{equation}
    \Delta \lambda = \alpha B
\end{equation}
gives $\alpha = 2.1$\,nm/mT and $5.1$\,nm/mT for LaFePO and CeCoIn$_5$ respectively. Note these are the raw numbers without any correction for demagnetising effects.  For  CeCoIn$_5$ any demagnetising effects are very small because of the $H\|ab$ geometry and for LaFePO ($H\|c$) the demagnetising effects cancel out as explained in the previous section.  As $\alpha = \lambda_0/H_0$ and taking $\lambda_0=240$\,nm for LaFePO \cite{Uemura2009} and $\lambda_0=190$\,nm for CeCoIn$_5$ \cite{Ormeno2002} this gives $\mu_0H_0=114$\,mT for LaFePO and $\mu_0H_0=36$\,mT for CeCoIn$_5$. 

Density function theory calculations of the electronic structure of  LaFePO show that there is considerable variation in $v_F$ on the different sheets and within each sheet of Fermi surface.  The location of the node(s) is presently unknown.  $v_F$ ranges from 60\,km/s to 500\,km/s with an average value of 280\,km/s (calculated from the DFT calculations described in Ref.\ \cite{Coldea2008,carrington2009}). de Haas van-Alphen measurements \cite{Coldea2008,carrington2009} show the effective mass is enhanced by a factor 2 compared to DFT and so $v_F$ is reduced by this factor.  Assuming   $\mu_\Delta=2$ and $\Delta_0=1.43T_c$ (values for single band $d$-wave), we calculate $\mu_0H_0$ ranging from 37\,mT to 308\,mT using Eq.\ (\ref{Eq:H0}).  For CeCoIn$_5$, the orbitally averaged $v_F$ derived from de Haas-van Alphen effect data \cite{Settai2001} ranges from 11\,km/s to 50\,km/s. With $T_c=2.1$\,K and again assuming a single band $d$-wave gap, this gives $\mu_0H_0$ in the range 80 to 370 \,mT.   

Anisotropy in $v_F$ can substantially affect $H_0$ as the non-linear energy shift depends on the value of $v_F$ at the node, but $\lambda_0$ is a weighted average over the whole Fermi surface \cite{Xu1995}. In materials, such as CeCoIn$_5$, where there are very strong mass renormalisations, there are additional complications regarding how these renormalisations affect the temperature dependent superconducting properties \cite{Miyake2018}.   Although the above estimated $H_0$ values are in the same range as the experimental ones, further theoretical work on multiband and strongly interacting superconductors is required before quantitative conclusions can be drawn.  In particular, the location of the nodes needs to be known.

\newpage
\bibliographystyle{apsrev}
\bibliography{../NLME2019}